\documentclass[12pt]{article}

\usepackage{cite, amssymb}
\usepackage{braket}

\expandafter\let\csname equation*\endcsname\relax
\expandafter\let\csname endequation*\endcsname\relax
\usepackage{amsmath}

\usepackage{array}
\usepackage{color}
\usepackage{physics}
\usepackage{tikz}
\usetikzlibrary{patterns}

\newcommand{\Hvier}{$H^{4}$}
\newcommand{\Hsechs}{$H^{6}$}
\newcommand{\Z}{\mathbb{Z}}

\newcommand{\N}{\mathbb{N}}

\DeclareMathOperator{\Id}{Id}

\newcommand{\tHeq}[2][]{${}_{t}H^{\varepsilon #1}_{#2}$}

\newcommand{\half}{\frac{1}{2}}
\renewcommand{\vec}{\mathbf}

\newcommand{\Fp}[1]{F^{(+)}_{#1}}
\newcommand{\Fm}[1]{F^{(-)}_{#1}}

\newcommand{\Fppp}{\Fp{n}\Fp{m}}
\newcommand{\Fpmm}{\Fp{n}\Fm{m}}
\newcommand{\Fmpm}{\Fm{n}\Fp{m}}
\newcommand{\Fmmp}{\Fm{n}\Fm{m}}
\newcounter{rmk}
\renewcommand{\thermk}{\arabic{rmk}}
\newenvironment{remark}%
{\refstepcounter{rmk}\vspace{6pt}\noindent\ignorespaces\textbf{Remark
\thermk:}}{\vspace{6pt}\par}
\renewcommand{\epsilon}{\varepsilon}
\renewcommand{\imath}{\mathrm{i}}

\makeatletter 
\renewcommand{\pdv}[2]{\begingroup 
  \@tempswafalse\toks@={}\count@=\z@ 
  \@for\next:=#2\do 
    {\expandafter\check@var\next\@nil
     \advance\count@\der@exp 
     \if@tempswa 
       \toks@=\expandafter{\the\toks@\,}%
     \else 
       \@tempswatrue 
     \fi 
     \toks@=\expandafter{\the\expandafter\toks@\expandafter\partial\der@var}}%
  \frac{\partial\ifnum\count@=\@ne\else^{\number\count@}\fi#1}{\the\toks@}%
  \endgroup} 
\def\check@var{\@ifstar{\mult@var}{\one@var}} 
\def\mult@var#1#2\@nil{\def\der@var{#2^{#1}}\def\der@exp{#1}} 
\def\one@var#1\@nil{\def\der@var{#1}\chardef\der@exp\@ne} 
\makeatother 

\allowdisplaybreaks

\title{Darboux integrability of trapezoidal $H^{4}$ and 
    $H^{6}$ families of lattice equations I:
    First integrals}
\author{G. Gubbiotti$^{1}$ and R.I. Yamilov$^2$}
\date{$^1$ Dipartimento di Matematica e Fisica, Universit\`a degli Studi Roma Tre
    and Sezione INFN di Roma Tre,
    Via della Vasca Navale 84, 00146 Roma, Italy
    \\
    E-mail: \texttt{gubbiotti@mat.uniroma3.it}
    \\
    $^2$ Institute of Mathematics, Ufa Scientific Center,
    Russian Academy of Sciences,
    112 Chernyshevsky Street, Ufa 450008, Russian Federation
    \\
    E-mail: \texttt{RvlYamilov@matem.anrb.ru}}

\begin{document}

\maketitle

\begin{abstract}
    In this paper we prove that the trapezoidal $H^{4}$
    and the $H^{6}$ families of quad-equations 
    are Darboux integrable
    systems. This result sheds light on the fact that such equations
    are linearizable as it was proved, using the Algebraic Entropy test
    [G. Gubbiotti, C. Scimiterna and D. Levi, Algebraic entropy, symmetries 
        and linearization for quad equations consistent on the cube,
    \emph{J. Nonlinear Math. Phys.}, 23(4):507–543, 2016].
    We conclude with some suggestions on how first integrals can
    be used to obtain general solutions.
\end{abstract}

\section{Introduction}

Since its introduction
the integrability criterion  denoted Consistency 
Around the Cube (CAC) has been a source of many results in the 
classification of nonlinear  {partial}
difference equations on a quad graph. 
 {The importance of this criterion is because it ensures
the existence of
B\"acklund transformations 
\cite{DoliwaSantini1997,Nijhoff2001,BobenkoSuris2002,Nijhoff2002,Bridgman2013} 
and, as a consequence, of Lax pairs.} As it is well known \cite{Yamilov2006},  Lax 
pairs and B\"acklund transforms are associated with both linearizable 
and integrable equations.  
 {We point out that to be \emph{bona fide}
a Lax pair has to give rise to a genuine
spectral problem \cite{CalogeroDeGasperisIST_I},
otherwise the Lax pair is called \emph{fake Lax pair}
\cite{CalogeroNucci1991,Hay2009,Hay2011,
HayButler2013,HayButler2015}.
A fake Lax pair is useless in proving (or disproving)
the integrability, since it can be equally found
for integrable and non-integrable
equations.
In the linearizable case
Lax pairs must be then fake ones, even if
proving it is usually a highly nontrivial
task \cite{GSL_Gallipoli15}.}

 {In \cite{ABS2003} was carried out
the first attempt to classify all the
multi-affine partial
difference equations defined on the quad
graph and possessing CAC.
In \cite{ABS2003} the quad graph was treated
as a geometric object not embedded in any
$\Z^{2}$-lattice, as displayed in Figure
\ref{fig:geomquad}.
Then the quad-equation is an expression
of the form:
\begin{equation}
    Q\left( x,x_{1},x_{2},x_{12};\alpha_{1},\alpha_{2}  \right) =0,
    \label{eq:quadequa}
\end{equation}
connecting some \emph{a priori} 
independent fields $x$, $x_{1}$, $x_{2}$, $x_{12}$
assigned to the vertices of the quad graph,
see Figure \ref{fig:geomquad}.
$Q$ is assumed to be
a \emph{multi-affine} polynomial
in $x$, $x_{1}$, $x_{2}$, $x_{12}$
and, as shown in Figure \ref{fig:geomquad},
$\alpha_{1}$ and $\alpha_{2}$ are parameters
assigned to the edges of the quad graph.

\begin{figure}[bthp]
   \centering
   \begin{tikzpicture}
       \node (x1) at (0,0) [circle,fill,label=-135:$x$] {};
       \node (x4) at (0,2.5) [circle,fill,label=135:$x_{1}$] {};
       \node (x2) at (2.5,0) [circle,fill,label=-45:$x_{2}$] {};
       \node (x3) at (2.5,2.5) [circle,fill,label=45:$x_{12}$] {};
       \draw [thick] (x2) to node[below] {$\alpha_{1}$} (x1);
       \draw [thick] (x4) to node[above] {$\alpha_{1}$} (x3);
       \draw [thick] (x3) to node[right] {$\alpha_{2}$} (x2);
       \draw [thick] (x1) to node[left] {$\alpha_{2}$} (x4);
   \end{tikzpicture} 
   \caption{ {The purely geometric quad graph not embedded 
   in any lattice.}}
\label{fig:geomquad}
\end{figure}
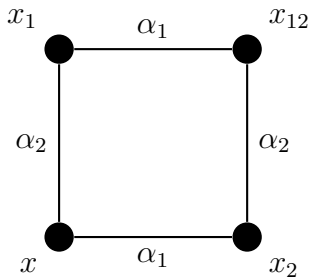

In this setting, we define the Consistency Around the Cube 
as follows: assume we are given given six quad-equations:
\begin{subequations}
    \begin{align}
        A\left(x,x_{1},x_{2},x_{12};\alpha_1, \alpha_2\right) &=0,
        \label{eq:Aeq}
        \\
        \bar{A}\left(x_{3},x_{13},x_{23},x_{123};\alpha_1, \alpha_2\right) &=0,
        \label{eq:Abeq}
        \\
        B\left(x,x_{2},x_{3},x_{23};\alpha_3, \alpha_2\right) &=0, 
        \label{eq:Beq}
        \\
        \bar{B}\left(x_{1},x_{12},x_{13},x_{123};\alpha_3, \alpha_2\right)&=0,
        \label{eq:Bbeq}
        \\
        C\left(x,x_{1},x_{3},x_{13};\alpha_1, \alpha_3\right) &=0,
        \label{eq:Ceq}
        \\
        \bar{C}\left(x_{2},x_{12},x_{23},x_{123};\alpha_1, \alpha_3\right) &=0,
        \label{eq:Cbeq}
    \end{align}
    \label{eq:system}
\end{subequations}
arranged on the faces of a cube
as in Figure \ref{fig:cube2}.
Then if $x_{123}$ computed from \eqref{eq:Abeq},
\eqref{eq:Bbeq} and \eqref{eq:Cbeq} coincide
we say that the system \eqref{eq:system}
possesses the Consistency Around the Cube.

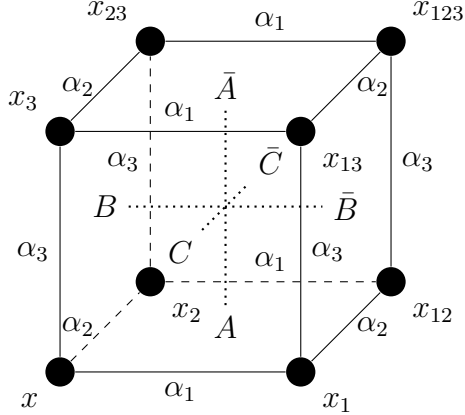
\begin{figure}[htbp] 
    \centering
    \begin{tikzpicture}[auto,scale=0.8]
        \node (x) at (0,0) [circle,fill,label=-135:$x$] {};
        \node (x1) at (4,0) [circle,fill,label=-45:$x_{1}$] {};
        \node (x2) at (1.5,1.5) [circle,fill,label=-45:$x_{2}$] {};
        \node (x3) at (0,4) [circle,fill,label=135:$x_{3}$] {};
        \node (x12) at (5.5,1.5) [circle,fill,label=-45:$x_{12}$] {};
        \node (x13) at (4,4) [circle,fill,label=-45:$x_{13}$] {};
        \node (x23) at (1.5,5.5) [circle,fill,label=135:$x_{23}$] {};
        \node (x123) at (5.5,5.5) [circle,fill,label=45:$x_{123}$] {};
        \node (A) at (2.75,0.75) {$A$};
        \node (Aq) at (2.75,4.75) {$\bar A$};
        \node (B) at (0.75,2.75) {$B$};
        \node (Bq) at (4.75,2.75) {$\bar B$};
        \node (C) at (2,2) {$C$};
        \node (Cq) at (3.5,3.5) {$\bar C$};
        \draw (x) -- node[below]{$\alpha_{1}$} (x1) 
        -- node[right] {$\alpha_{2}$} (x12) -- node[right] {$\alpha_{3}$} (x123) 
        -- node[above] {$\alpha_{1}$} (x23) -- node[left] {$\alpha_{2}$} (x3) -- node[left] {$\alpha_{3}$} (x);
        \draw (x3) -- node[above]{$\alpha_{1}$} (x13) -- node[right]{$\alpha_{3}$} (x1);
        \draw (x13) -- node[right]{$\alpha_{2}$} (x123);
        \draw [dashed] (x) --node[left]{$\alpha_{2}$} (x2) 
        -- node[above] {$\alpha_{1}$} (x12);
        \draw [dashed] (x2) --node[left] {$\alpha_{3}$} (x23);
        \draw [dotted,thick] (A) to (Aq);
        \draw [dotted,thick] (B) to (Bq);
        \draw [dotted,thick] (C) to (Cq);
    \end{tikzpicture} 
    \caption{ {Equations on a Cube}}
    \label{fig:cube2}
\end{figure}

In \cite{ABS2003} the classification
was carried out up to the action of
a general M\"obius transformation and up to
point transformations of the edge parameters,
with the additional assumptions:
\begin{enumerate}
    \item All the faces of the cube in Figure
        \ref{fig:cube2} carry the same equation
        up to the edge parameters.
    \item The quad-equation \eqref{eq:quadequa} 
        possesses the
        $D_{4}$ discrete symmetries:
        \begin{equation}
            \begin{aligned}
                Q\left( x,x_{1},x_{2},x_{12};\alpha_{1},\alpha_{2} \right)
                &= \mu
                Q\left( x,x_{2},x_{1},x_{12};\alpha_{2},\alpha_{1} \right)
                \\
                &= \mu'
                Q\left( x_{1},x,x_{12},x_{2};\alpha_{1},\alpha_{2} \right),
            \end{aligned}
            \label{eq:squaresymm}
        \end{equation}
        where $\mu,\mu'\in\Set{\pm1}$.
    \item The system \eqref{eq:system} possesses
        the \emph{tetrahedron property}, i.e.
        $x_{123}$ is independent of $x$:
        \begin{equation}
            x_{123} = 
            x_{123}\left( x,x_{1},x_{2},x_{3};\alpha_{1},\alpha_{2},\alpha_{3} \right) 
            \implies
            \pdv{x_{123}}{x} = 0.
            \label{eq:tetrahedron}
        \end{equation}
\end{enumerate}
}
The result was then two classes of discrete autonomous 
equations {: the $H$ and $Q$ equations}.

Releasing the hypothesis that every face of the cube carried
the same equation, the same authors in \cite{ABS2009} presented some
new equations  {without classification purposes}.

A complete classification in this extended setting
was then accomplished by R. Boll in a series of papers
culminating in his Ph.D. thesis \cite{Boll2011,Boll2012a,Boll2012b}.
In these papers the classification of all
the  {consistent} sextuples of partial difference equations
on the quad graph,
 {i.e. systems of the form \eqref{eq:system}}, has been carried out.
 {The only technical assumption used in
\cite{Boll2011,Boll2012a,Boll2012b} is
the tetrahedron property.}
The obtained equations
may fall into three disjoint families
depending on their bi-quadratics:
\begin{equation}
    h_{ij}=\pdv{Q}{y_{k}}\pdv{Q}{y_{l}}-Q\pdv{Q}{y_{k},y_{l}}, \qquad 
		Q=Q\left( y_{1},y_{2},y_{3},y_{4};\alpha_1,\alpha_2  \right),
    \label{eq:biquadr}
\end{equation}
where we use a special notation for variables of $Q$, and the pair $\{k,l\}$ is the complement of the pair $\{i,j\}$
in $\left\{ 1,2,3,4 \right\}$. 
A bi-quadratic is called
\emph{degenerate} if it contains linear factors of the form $y_{i}-c$,
 {where $c$ is a constant}, otherwise a bi-quadratic is called \emph{non-degenerate}.
The three families are classified depending on how many bi-quadratics
are degenerate:
\begin{itemize}
    \item $Q$-type equations: all the bi-quadratics are nondegenerate,
    \item $H^{4}$-type equations: four bi-quadratics are degenerate,
    \item $H^{6}$-type equations: all of the six bi-quadratics are degenerate.
\end{itemize}
Let us notice that the $Q$ family is the same as that which was introduced 
in \cite{ABS2003}.
The $H^{4}$ equations are divided into two subclasses: \emph{rhombic}
and \emph{trapezoidal}, depending on their discrete symmetries.

 {We remark that all classification results hold locally 
in the sense that they relate to a single quadrilateral
cell or a single cube displayed in Figures
\ref{fig:geomquad} and \ref{fig:cube2}.
The important problem of embedding these results into 
a two- or three-dimensional lattice, with preservation of the three-dimensional consistency 
condition, was already discussed in \cite{ABS2009,Xenitidis2009}
by using the concept of a Black and White lattice.
One way to solve this problem is
to embed \eqref{eq:quadequa} into a $\Z^2$-lattice with an elementary cell of size greater 
than one. In this case, the quad-equation 
\eqref{eq:quadequa} can be extended to a lattice, 
and the lattice equation becomes integrable or
linearizable. To this end, following 
\cite{Boll2011,Boll2012a,Boll2012b}, 
we reflect the square with respect to the normal 
to its right and top sides and then complete a 
$2\times2$ lattice by again reflecting one of the 
obtained squares in the other direction. 
Such procedure is graphically described in Figure
\ref{fig:elcell}.}

 {
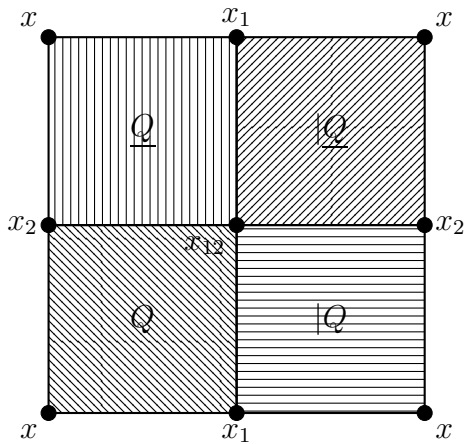
\begin{figure}[htpb]
\centering
\begin{tikzpicture}[scale=2.5]
    \draw [pattern=north west lines,thick] (0,0) rectangle (1,1);
    \draw [pattern=north east lines,thick] (1,1) rectangle (2,2);
    \draw [pattern=horizontal lines,thick] (1,0) rectangle (2,1);
    \draw [pattern=vertical lines,thick] (0,1) rectangle (1,2);
    \foreach \x in {0,...,2}{
        \foreach \y in {0,...,2}{
            \node[draw,circle,inner sep=2pt,fill] at (\x,\y) {};
        }
    }
    \node[below left] at (0,0) {$x$};
    \node[below] at (1,0) {$x_1$};
    \node[below right] at (2,0) {$x$};
    \node[left] at (0,1) {$x_2$};
    \node[below left] at (1,1) {$x_{12}$};
    \node[right] at (2,1) {$x_2$};
    \node[above left] at (0,2) {$x$};
    \node[above] at (1,2) {$x_1$};
    \node[above right] at (2,2) {$x$};
    \node[] at (1/2,1/2) {$Q$};
    \node[] at (1/2+1,1/2) {$|Q$};
    \node[] at (1/2,1/2+1) {$\underline{Q}$};
    \node[] at (1/2+1,1/2+1) {$|\underline{Q}$};
\end{tikzpicture}
\caption{ {The ``four stripe'' lattice.}}
\label{fig:elcell}
\end{figure}
}

 {This corresponds to constructing three  equations obtained 
from \eqref{eq:quadequa} by flipping its arguments:
\begin{subequations}
    \begin{align}
        Q &=Q(x,x_{1},x_{2},x_{12};\alpha_{1},\alpha_{2}) =0,
        \\
        |Q &=Q(x_{1},x,x_{12},x_{2};\alpha_{1},\alpha_{2}) =0,
        \\
        \underline{Q} &= Q(x_{2},x_{12},x,x_{1};\alpha_{1},\alpha_{2}) =0,
        \\
        |\underline{Q} &= Q(x_{12},x_{2},x_{1},x;\alpha_{1},\alpha_{2}) =0.
    \end{align}
\label{eqn:dysys1}%
\end{subequations}
By paving the whole $\Z^{2}$ with such equations, we 
get a partial difference equation which can be in 
principle studied using known methods. 
Since \emph{a priori}  
$Q\neq |Q \neq \underline{Q} \neq |\underline{Q}$,  
the obtained lattice will be
a four stripe lattice, i.e. an extension of the Black and White 
lattice considered, for example, in 
\cite{ABS2009,Xenitidis2009,HietarintaViallet2012}.
This gives rise to lattice equations with two-periodic
coefficients for an unknown function $u_{n,m}$, with $(n,m)\in\Z^{2}$:
\begin{equation}
    \begin{aligned}
        &\phantom{+}\Fppp Q(u_{n,m},u_{n+1,m},u_{n,m+1},u_{n+1,m+1};\alpha_{1},\alpha_{2})
    \\
    &+\Fmpm|Q (u_{n,m},u_{n+1,m},u_{n,m+1},u_{n+1,m+1};\alpha_{1},\alpha_{2})
    \\
    &+\Fpmm\underline{Q} (u_{n,m},u_{n+1,m},u_{n,m+1},u_{n+1,m+1};\alpha_{1},\alpha_{2})
    \\
    &+\Fmmp|\underline{Q} (u_{n,m},u_{n+1,m},u_{n,m+1},u_{n+1,m+1};\alpha_{1},\alpha_{2})
    =0,
    \end{aligned}
    \label{eq:dysys3}
\end{equation}
where
\begin{equation}
    F_{k}^{(\pm)} = \frac{1\pm\left( -1 \right)^{k}}{2}.
    \label{eq:fk}
\end{equation}
This explicit formula was first presented
in \cite{GSL_Gallipoli15}.
}
For more details on the construction of equations on the lattice
from the single cell equations,
we refer to  {\cite{Xenitidis2009,Boll2011,Boll2012a,Boll2012b}}
and to the Appendix in \cite{GSL_general}.

A detailed study of all the lattice equations derived 
from the \emph{rhombic} $H^{4}$ 
family, including the construction of their three-leg forms, 
Lax pairs, B\"acklund transformations and infinite hierarchies 
of generalized symmetries, 
has been presented in \cite{Xenitidis2009}.
 {On the other hand} in \cite{GSL_general}, 
by calculating the Algebraic Entropy 
\cite{BellonViallet1999,Viallet2006,HietarintaViallet2007,Viallet2009},
it was suggested that the \emph{trapezoidal \Hvier
and the \Hsechs~equations were linearizable}.
It has been proved that the  {rate of} growth 
of the equations belonging
to these two families is \emph{linear}, that implies the linearizability,
and two explicit examples of linearization have been given.
In \cite{GSL_Gallipoli15} a particular example,
the \tHeq{1} equation, was studied and it was found that it possessed three-point
generalized symmetries depending on arbitrary functions. This property
was later linked in \cite{GSL_Pavel} to the fact that the
\tHeq{1} was \emph{Darboux integrable}.
In addition in \cite{GSL_Pavel} it was proved that some other
quad-equations Consistent Around the Cube, which were known to be
linearizable \cite{Hietarinta2004,Hietarinta2005}, were in fact
Darboux integrable.
These facts provide some evidence of an intimate connection between
linearizable equations with CAC and Darboux integrability.

The scope of this paper is to generalize the result obtained for
the \tHeq{1} in \cite{GSL_Gallipoli15,GSL_Pavel}.
Our main statement is the following:
\begin{quote}
    \emph{Every trapezoidal \Hvier~equation and every \Hsechs~equation
        presented in \cite{Boll2011,Boll2012a,Boll2012b}
    is Darboux integrable}. 
\end{quote}
The fact that an equation is Darboux
integrable is a \emph{formal} proof that it is linearizable,
as it will be discussed in more detail in Section \ref{sec:darboux}.

The plan of the paper is the following: in Section 
\ref{sec:darboux} we recall the basic facts about Darboux 
integrability and discuss the methodologies employed in the
case of non-autonomous, two-periodic quad equations. 
In Section \ref{sec:fint}
we present the first integrals of the trapezoidal \Hvier~
equations and of the \Hsechs~equations. In the Section \ref{sec:concl}
we present some conclusions and provide an outlook on how
first integrals can be used to obtain general solutions.

\section{Darboux integrability}
\label{sec:darboux}

In the continuous case, a hyperbolic partial differential
equation (PDE) in two variables
\begin{equation}
    u_{xt} = f\left(x,t, u,u_{t},u_{x} \right)
    \label{eq:hyppde}
\end{equation}
is said to be \emph{Darboux integrable} if it possesses two
independent first integrals $T,X$  depending only on
derivatives with respect to one variable:
\begin{equation}
    \begin{aligned}
        T = T\left(x,t,u,u_{t},\dots,u_{nt} \right), &\quad  
        \left.\dv{T}{x}\right|_{u_{xt}=f} \equiv0,
        \\
        X = X\left(x,t,u,u_{x},\dots,u_{mx} \right), &\quad 
        \left.\dv{X}{t}\right|_{u_{xt}=f} \equiv0,
    \end{aligned}
    \label{eq:hyppdeint}
\end{equation}
where $u_{kt}=\partial^{k} u / \partial t^{k}$ and 
$u_{kx}=\partial^{k} u / \partial x^{k}$ for every $k\in \N$.
The method is based on the linear theory
developed by Euler and Laplace \cite{Euler1768_III,Laplace1773}
and extended to the nonlinear case in the  {19th} 
and early  {20th} centuries
\cite{Darboux1870,Darboux1915,Goursat1896,Gau1911,Vessiot1939}.
The method was then used at the end of the  {20th} century mainly 
by Russian mathematicians as a source of new exactly solvable PDEs in two variables 
\cite{ZhiberIbragimovShabat1979,ZhiberShabat1979,ZhiberShabat1984,
    ZhiberSokolovStartsev1995,ZhiberSokolov1995,
    ZhiberSokolov1999,ZhiberSokolov2011}.
We note that in many papers  {Darboux integrability
    is defined as the stabilization to zero of the 
so-called Laplace chain of the linearized equation.}
However it can be proved that the two definitions are 
equivalent, to this end see \cite{AndersonKamran1994,AndersonKamran1997,
    AndersonJuras1997,ZhiberSokolovStartsev1995}.

The most famous equation, which belongs to the class
of Darboux integrable equations, is the Liouville
equation \cite{Liouville1853}:
\begin{equation}
    u_{xt} = e^{u}
    \label{eq:liouville}
\end{equation}
which possesses the two following first integrals:
\begin{equation}
     X =  u_{xx} - \half u_{x}^{2}, 
     \quad 
     T =  u_{tt} - \half u_{t}^{2}.
    \label{eq:liouvilleint}
\end{equation}

In the discrete setting Darboux integrability was introduced 
in \cite{AdlerStartsev1999},  {where it was} used to obtain
a discrete analogue of the Liouville equation \eqref{eq:liouville}.
Similarly as in the continuous case, we 
say that a quad-equation, possibly non-autonomous: 
\begin{equation}
    Q_{n,m}\left( u_{n,m},u_{n+1,m},u_{n,m+1},u_{n+1,m+1} \right)=0,
    \label{eq:quadequana}
\end{equation}
is \emph{Darboux integrable} if there exist two independent
first integrals, one containing only shifts in the first
direction and the other containing only shifts in the second direction.
This means that there exist two functions:
 {
\begin{subequations}
    \begin{align}
        W_1=W_{1,n,m}(u_{n+l_1,m},u_{n+l_1+1,m},\ldots,u_{n+k_1,m}),
        \label{eq:darbfint1}
        \\
        W_2=W_{2,n,m}(u_{n,m+l_2},u_{n,m+l_2+1},\ldots,u_{n,m+k_2}),
        \label{eq:darbfint2}
    \end{align}
    \label{eq:darbfint}
\end{subequations}
}
where $l_1<k_1$ and $l_2<k_2$ are integers, such that the relations
 {
\begin{subequations}
    \begin{align}
        (T_n-\Id)W_2=0, 
        \label{eq:darb1}
        \\
        (T_m-\Id)W_1=0
        \label{eq:darb2}
    \end{align}
    \label{eq:darbdef}
\end{subequations}
}
 {hold true identically on the solutions of
    \eqref{eq:quadequana}}.
By $T_n,T_m$ we denote the shift operators 
in the first and second 
directions, i.e. $T_n h_{n,m}=h_{n+1,m}$, $T_m h_{n,m}=h_{n,m+1}$,
 {and} by $\Id$ we denote the identity 
operator $\Id h_{n,m}=h_{n,m}$.
 {
The number $k_{i}-l_{i}$, where
$i=1,2$, is called the \emph{order} of
the first integral $W_{i}$.}

 {We notice that the existence of
first integrals implies that the following two transformations: 
\begin{subequations}
    \begin{align}
        u_{n,m} &\to \tilde u_{n,m} = W_{1,n,m},
        \\
        u_{n,m} &\to \hat u_{n,m} = W_{2,n,m}
    \end{align}
    \label{eq:fintlin}
\end{subequations}
bring the quad-equation \eqref{eq:quadequana}
into \emph{trivial linear equations} given by 
\eqref{eq:darbdef} \cite{AdlerStartsev1999}, namely:
\begin{subequations}
    \begin{align}
        \tilde u_{n,m+1} - \tilde u_{n,m} = 0 , 
        \\  
        \hat u_{n+1,m} - \hat u_{n,m} = 0.
    \end{align}
    \label{eq:fintlin2}
\end{subequations}
Therefore any Darboux integrable equation is linearizable in
two different ways.
This is the relationship between the Darboux integrability 
and linearization. 

The transformations \eqref{eq:fintlin} along
with the relations \eqref{eq:fintlin2} imply 
the following relations:
\begin{subequations}
    \begin{align}
        W_{1,n,m}=\lambda_{n},
        \label{eq:darbeq1}
        \\
        W_{2,n,m} =\rho_{m},
        \label{eq:darbeq2}
    \end{align}
    \label{eq:darbeq}
\end{subequations}
where $\lambda_{n}$ and $\rho_{m}$ are
arbitrary functions of the lattice variables
$n$ and $m$, respectively.
The relations \eqref{eq:darbeq} can
be seen as \emph{ordinary difference
equations} which must be satisfied
by any solution $u_{n,m}$ of \eqref{eq:quadequana}.
The transformations \eqref{eq:fintlin} 
and the ordinary difference
equations \eqref{eq:darbeq} may be
quite complicated. However in case of the
trapezoidal
\Hvier\ and the \Hsechs\ equations, we will prove
in a future work \cite{GSY_DarbouxII} that
also the equations \eqref{eq:darbeq}, defined by the first
integrals, are \emph{linearizable}. Therefore
we can use Darboux integrability in order
to obtain the \emph{general solutions} of these
equations.
In Section \ref{sec:concl} we will
present an example of this procedure
for the $_{t}H_{1}^{\varepsilon}$ equation.
}

 {From its introduction in \cite{AdlerStartsev1999},
    various papers were devoted to the
study of Darboux integrability for quad-equations
\cite{Habibullin2005,Habibullinetal2011,
    GarifullinYamilov2012,Startsev2014,
    GarifullinYamilov2015}.
In particular in \cite{Habibullin2005,GarifullinYamilov2012,
GarifullinYamilov2015} were developed
computational methods  to compute the first integrals.}
 {In \cite{Habibullin2005} was presented} a method
to compute the first integrals with fixed $l_{i}$, $k_{i}$ of
a given autonomous equation.
 {Then in} \cite{GarifullinYamilov2012} this method 
was slightly modified and applied to autonomous equations
with non-autonomous first integrals. 
 {Finally} in \cite{GarifullinYamilov2015}
it was applied to equations with two-periodic coefficients.
In the present paper, we present a further modification 
of this method 
for dealing with  {non-autonomous equations
with two-periodic coefficients}.

If we consider the operator
\begin{equation}
    Y_{-1} = T_{m} \pdv{}{u_{n,m-1}}T_{m}^{-1}
    \label{eq:ym1}
\end{equation}
and apply it to the definition of first integral in the $n$-direction
\eqref{eq:darb2}, we obtain:
\begin{equation}
    Y_{-1} W_{1} \equiv 0.
    \label{eq:ym1w1}
\end{equation}
The application of the operator $Y_{-1}$ is to be understood
in the following sense:
first we must apply $T_m^{-1}$ and then we should express, 
using the equation \eqref{eq:quadequana}, 
$u_{n+i,m-1}$ in terms of the functions $u_{n+j,m}$ and $u_{n,m-1}$ 
which will be considered in this problem as independent variables. 
Then we can differentiate in \eqref{eq:ym1w1}
with respect to $u_{n,m-1}$
and safely apply $T_{m}$ \cite{GarifullinYamilov2012}. 

Taking in \eqref{eq:ym1w1} the coefficients at powers of  $u_{n,m+1}$,
we obtain a system of PDEs for $W_{1}$. If this is sufficient to determine $W_{1}$ up to arbitrary 
functions of a single variable, then we are done, otherwise we can add
similar equations by considering the ``higher-order'' operators
\begin{equation}
    Y_{-k} = T_{m}^{k} \pdv{}{u_{n,m-1}}T_{m}^{-k}, \quad k\in \N,
    \label{eq:ymk}
\end{equation}
which annihilate the 
 {difference consequence} of \eqref{eq:darb2} given
by $T_{m}^{k}W_{1} = W_{1}$:
\begin{equation}
    Y_{-k} W_{1} \equiv 0, \quad k\in\N,
    \label{eq:ymkw1}
\end{equation}
with the same computational prescriptions as above. 
We can  {add equations
until we find a non-constant function\footnote{Obviously 
constant functions are trivial first integrals.} $W_1$ 
which depends on a \emph{single} combination of the variables
$u_{n,m+j_{1}}$, \dots, $u_{n,m+k_{1}}$. }
 {
If we find a non-constant solution $W_1$ of the equations generated
by \eqref{eq:ym1} and possibly \eqref{eq:ymk}, then we must insert it
into \eqref{eq:darb2} to specify it.
}

In the same way  {first integrals in the $m$-direction}
$W_{2}$ can be found by considering the
operators
\begin{equation}
    Z_{-k} = T_{n}^{k} \pdv{}{u_{n-1,m}}T_{n}^{-k}, \quad k\in\N ,
    \label{eq:zmk}
\end{equation}
which provide the equations
\begin{equation}
    Z_{-k} W_{2} \equiv 0, \quad k\in\N.
    \label{eq:zmkw2}
\end{equation}

In the case of non-autonomous equations with two-periodic coefficients,
we can assume  {that a decomposition analogue of
    the quad-equation \eqref{eq:dysys3} holds for the first integrals}:
\begin{equation}
    \begin{aligned}
        W_{i} &= \Fppp W_{i}^{(+,+)}+\Fmpm W_{i}^{(-,+)}
        \\
        &+\Fpmm W_{i}^{(+,-)} + \Fmmp W_{i}^{(-,-)},
    \end{aligned}
    \label{eq:widec}
\end{equation}
with  {$F_{k}^{(\pm)}$ given by \eqref{eq:fk}}. 
 {We can then} 
derive from (\ref{eq:ymkw1},\ref{eq:zmkw2}) a set of equations
for the functions $W_{i}^{(\pm,\pm)}$ by considering the even/odd
points on the lattice. The final form of the functions $W_{i}$ will be
then fixed by substituting in \eqref{eq:darbdef} and
separating again.

As an example let us consider the problem of finding the
first integrals of the \tHeq{1} equation which was solved in \cite{GSL_Pavel} by
direct inspection:
\begin{equation}
    _{t}H_{1}^{\varepsilon}\colon 
        \begin{aligned}[t]
        &\left(u_{n,m}-u_{n+1,m}\right)  \left(u_{n,m+1}-u_{n+1,m+1}\right)
        \\ 
        &-\alpha_{2}\epsilon^2\left({F}_{m}^{\left(+\right)}u_{n,m+1}u_{n+1,m+1}
        +{F}_{m}^{\left(-\right)}u_{n,m}u_{n+1,m}\right) -\alpha_{2}=0.
        \end{aligned}
    \label{eq:tH1e}
\end{equation}
Since all the \Hvier~equations and the \tHeq{1}, in 
particular, are non-autonomous only in the direction $m$, 
we can consider a simplified version of \eqref{eq:widec}:
\begin{equation}
    W_{i} = \Fp{m} W_{i}^{(+)}+\Fm{m} W_{i}^{(-)}.
    \label{eq:widecsimp}
\end{equation}

If we assume that $W_{1}=W_{1,n,m}\left( u_{n,m},u_{n+1,m} \right)$,  
then,
separating the even and odd terms with respect 
to $m$ in \eqref{eq:ym1w1},
we find the following equations:
\begin{subequations}
    \begin{align}
        \frac {\partial W_{1}^{(+)}}{\partial u_{{n+1,m}}} 
        +\frac {\partial W_{1}^{(+)}}{\partial u_{{n,m}}} &= 0,
        \\
        \left(1+{\epsilon}^{2}u_{{n+1,m}}^{2}\right) 
        \frac {\partial W_{1}^{(-)}}{\partial u_{{n+1,m}}}
        +\left(1+{\epsilon}^{2}u_{{n,m}}^{2}\right) 
        \frac {\partial W_{1}^{(-)}}{\partial u_{{n,m}}}
         &=0.
    \end{align}
    \label{eq:ym1w1tH1e}
\end{subequations}
Their solution is:
\begin{equation}
    W_{1} = 
    \Fp{m} F\left(u_{n+1, m}-u_{n, m}\right)
    +\Fm{m} G\left(\frac{u_{n+1, m}-u_{n, m}}{1+\epsilon^2 u_{n, m}u_{n+1, m}}\right) ,
    \label{eq:W1tH1e1}
\end{equation}
where $F$ and $G$ are arbitrary functions.
Inserting \eqref{eq:W1tH1e1} into the difference equation \eqref{eq:darb2},
we obtain that $F$ and $G$ must satisfy the following identity:
\begin{equation}
    G\left( \xi \right) = F\left( \frac{\alpha_{2}}{\xi} \right).
    \label{eq:FGid}
\end{equation}
This yields the first integral
\begin{equation}
    W_{1} = \Fp{m}F\left(\frac{\alpha_{2}}{u_{n+1,m}-u_{n,m}}\right)
    +\Fm{m}F\left( \frac{u_{n+1,m}-u_{n,m}}{1+\varepsilon^{2}u_{n,m}u_{n+1,m}} \right).
    \label{eq:W1tH1e}
\end{equation}

For the $m$-direction we may also suppose that our first
integral $W_{2}=W_{2,m}\left( u_{n,m},u_{n,m+1} \right)$ is of the first order or 
a \emph{two-point} first integral.
It easy to see from \eqref{eq:zmkw2} with $k=1$ that this yields only the trivial
solution $W_{2}=\text{constant}$. Therefore we consider the case of a second order, 
\emph{three-point} first integral:
$W_{2}=W_{2,m}\left( u_{n,m-1},u_{n,m},u_{n,m+1} \right)$. 
From \eqref{eq:zmkw2} with $k=1$, separating the even and odd terms with respect to $m$, we obtain:
\begin{subequations}
    \begin{align}
        \begin{aligned}
            \alpha_{2}\left(1+ {\epsilon}^{2}u_{n,m+1}^{2}\right) 
            \frac {\partial W_{2}^{(+)}}{\partial u_{n,m+1}}
            &- \left[ \left( u_{n,m}-u_{n+1,m} \right)^{2}+\epsilon^{2}\alpha_{2}^{2} \right]
            {\frac {\partial W_{2}^{(+)}}{\partial u_{{n,m}}}}
            \\
            &+ \alpha_{2}\left(1+ \epsilon^{2}u_{{n,m-1}}^{2}\right) 
            \frac {\partial W_{2}^{(+)}}{\partial u_{{n,m-1}}}=0,
        \end{aligned}
        \\
        \begin{aligned}
            \alpha_{2}\left(1+ {\epsilon}^{2}u_{n+1,m}^{2}\right) 
            \frac {\partial W_{2}^{(-)}}{\partial u_{n,m+1}}
            &- \left( u_{n,m}-u_{n+1,m} \right)^{2}
            {\frac {\partial W_{2}^{(-)}}{\partial u_{{n,m}}}}
            \\
            &+ \alpha_{2}\left(1+ \epsilon^{2}u_{n+1,m}^{2}\right) 
            \frac {\partial W_{2}^{(-)}}{\partial u_{{n,m-1}}}=0.
        \end{aligned}
    \end{align}
    \label{eq:z1m1w1tH1e}
\end{subequations}
Taking the coefficients with respect to $u_{n+1,m}$ and then solving,
we have:
\begin{equation}
        W_{2} = \Fp{m} F\left( \frac{1+\varepsilon^{2}u_{n,m+1}u_{n,m-1}}{%
            u_{n,m+1}-u_{n,m-1}} \right)
            +\Fm{m} G\left( u_{n,m+1}-u_{n,m-1} \right).
        \label{eq:W2tH1e}
\end{equation}
Inserting \eqref{eq:W2tH1e} into \eqref{eq:darb1} we do not have 
any further restriction on the form
of the first integral. So we conclude that we have two independent
first integrals in the $m$-direction, as it was observed in \cite{GSL_Pavel}.

The fact that, when successful, the above procedure
gives arbitrary functions has to be understood as a restatement of the
trivial property that any autonomous function of a first integral is again a first
integral.
So, in general, one does not need first integrals depending on arbitrary
functions.
 {
Therefore we can take these arbitrary functions in
the first integrals to be \emph{linear} function in their arguments.} 


With these simplifying assumptions we can consider 
the first integrals for the \tHeq{1}  {equation} as given by:
\begin{subequations}
    \begin{align}
        W_{1} &= \Fp{m}\frac{\alpha_{2}}{u_{n+1,m}-u_{n,m}}
        +\Fm{m}\frac{u_{n+1,m}-u_{n,m}}{1+\varepsilon^{2}u_{n,m}u_{n+1,m}},
        \label{eq:W1tH1elin}
        \\
        W_{2} &= \Fp{m} \alpha \frac{
            1+\varepsilon^{2}u_{n,m+1}u_{n,m-1}}{u_{n,m+1}-u_{n,m-1}}
            +\Fm{m} \beta\left( u_{n,m+1}-u_{n,m-1} \right),
        \label{eq:W2tH1elin}
    \end{align}
    \label{eq:tH1eintegralslin}
\end{subequations}
where $\alpha$ and $\beta$ are two arbitrary constants.
This was the form in which the first integrals for the
\tHeq{1} equation \eqref{eq:tH1e} were presented in \cite{GSL_Pavel}.
In what follows we will write down the first integrals
according to  {above} prescription.

In next Section we present the explicit form of first integrals
for the remaining $H^{4}$ and $H^{6}$ equations computed by
the method presented in this Section. We will not present the
details of the calculations, since they are algorithmic and they
can be implemented 
in any Computer Algebra
System available (we have implemented them in \texttt{Maple}).

\section{First integrals for the \Hvier~and \Hsechs~equations} 
\label{sec:fint}

In this Section we consider the \tHeq{2}, \tHeq{3} equations
and the whole family of the $H^{6}$ equations.
As we stated in Introduction, these equations belong to 
the classification of Boll \cite{ABS2009,Boll2011,Boll2012a,Boll2012b}, 
and we will consider their non-autonomous form as
given in \cite{GSL_general}. 

\subsection{Trapezoidal \Hvier~equations}

We consider the trapezoidal \tHeq{2}, \tHeq{3} equations
which, in non-autonomous form \cite{GSL_general}, read as:
\begin{subequations}
    \begin{align}
        _{t}H_{2}\colon &
        \begin{aligned}[t]
        &\left(u_{n,m}-u_{n+1,m}\right)\left(u_{n,m+1}-u_{n+1,m+1}\right)
        \\
        &+\alpha_{2}\left(u_{n,m}+u_{n+1,m}+u_{n,m+1}+u_{n+1,m+1}\right)
        \\
        &+\frac{\epsilon\alpha_{2}}{2} \left(2{F}_{m}^{\left(+\right)}u_{n,m+1}
        +2\alpha_{3}+\alpha_{2}\right)\left(2{F}_{m}^{\left(+\right)}u_{n+1,m+1}+2\alpha_{3}+\alpha_{2}\right)
        \\
        &+\frac{\epsilon\alpha_{2}}{2} \left(2{F}_{m}^{\left(-\right)}u_{n,m}+2\alpha_{3}
        +\alpha_{2}\right)\left(2{F}_{m}^{\left(-\right)}u_{n+1,m}+2\alpha_{3}+\alpha_{2}\right)
        \\
        &+\left(\alpha_{3}+\alpha_{2}\right)^2-\alpha_{3}^2-2\epsilon\alpha_{2}\alpha_{3}\left(\alpha_{3}+\alpha_{2}\right)=0 ,
        \end{aligned}
        \label{eq:tH2e}
        \\
        _{t}H_{3}\colon &
        \begin{aligned}[t]
        &\alpha_{2}\left(u_{n,m}u_{n+1,m+1}+u_{n+1,m}u_{n,m+1}\right)
        \\
        &-\left(u_{n,m}u_{n,m+1}+u_{n+1,m}u_{n+1,m+1}\right)
        -\alpha_{3}\left(\alpha_{2}^{2}-1\right)\delta^2
        \\
        &-\frac{\epsilon^2(\alpha_{2}^{2}-1)}{\alpha_{3}\alpha_{2}}
        \left({{F}_{m}^{\left(+\right)}u_{n,m+1}u_{n+1,m+1}
        +{F}_{m}^{\left(-\right)}u_{n,m}u_{n+1,m}}\right)=0,
        \end{aligned}
        \label{eq:tH3e}
    \end{align}
    \label{eq:trapezoidalH4}
\end{subequations}
where $F_{m}^{(\pm)}$ are defined by \eqref{eq:fk}.

We now present the first integrals of \eqref{eq:trapezoidalH4} 
in both directions.

\subsubsection{\tHeq{2} equation}

Let us consider the \tHeq{2} equation as it is given by \eqref{eq:tH2e}. It has a
four-point, third order first integral in the $n$-direction:
\begin{subequations}
    \begin{equation}
        W_{1}  
        \begin{aligned}[t]
            &=\Fp{m}{\frac { \left( -u_{{n+1,m}}+u_{{n-1,m}} \right)  
            \left( u_{{n,m}}-u_{{n+2,m}} \right) }{
                {\epsilon}^{2}\alpha_{2}^{4}+4\epsilon\alpha_{2}^{3}+ 
                \left[  \left(8{ \alpha_{3}} -2u_{{n,m}}-2u_{{n+1,m}} \right) \epsilon-1 \right]
                \alpha_{2}^{2}+\left( u_{{n,m}}-u_{{n+1,m}} \right) ^{2}}}
            \\
            &-\Fm{m}{\frac { \left( -u_{{n+1,m}}+u_{{n-1,m}} \right)  
            \left( u_{{n,m}}-u_{{n+2,m}} \right) }{ \left( -u_{{n-1,m}}+u_{{n,m}}+{ \alpha_{2}} \right)  
            \left( u_{{n+1,m}}+{ \alpha_{2}}-u_{{n+2,m}} \right)}}
        \end{aligned}
        \label{eq:W1tH2e}
    \end{equation}
    and a five-point, fourth order first integral in the $m$-direction:
    \begin{equation}
        W_{2} 
        \begin{aligned}[t]
            &=\Fp{m}\alpha {\frac { \left( u_{{n,m-1}}-u_{{n,m+1}}\right) ^{2} 
            \left( u_{{n,m+2}}-u_{{n,m}} \right)  \left( u_{{n,m}} -u_{{n,m-2}}\right) }{ 
                \begin{gathered}
                \left(  \left( { \alpha_{2}}+{ \alpha_{3}}+u_{{n,m-1}} \right) ^{2}\epsilon
                -u_{{n,m-1}}+{ \alpha_{3}}-u_{{n,m}} \right)  \cdot
                \\
                \left(  \left( { \alpha_{3}}+{ \alpha_{2}}+u_{{n,m+1}} \right) ^{2}\epsilon
                -u_{{n,m+1}}+{ \alpha_{3}}-u_{{n,m}} \right) 
                \end{gathered}}}
            \\
            &+\Fm{m}\beta{\frac {\left[
                \begin{gathered}
                -\epsilon \left( u_{{n,m-2}}-u_{{n,m+2}} \right) 
            u_{n,m}^{2}- \left( { \alpha_{3}}+{ \alpha_{2}} \right) ^{2}
            \left( u_{{n,m-2}}-u_{{n,m+2}} \right) \epsilon
            \\
            +\left( -2 \left( u_{{n,m-2}}-u_{{n,m+2}} \right)  
            \left( { \alpha_{3}}+{ \alpha_{2}} \right) \epsilon+u_{{n,m-1}}
            -u_{{n,m+2}}-u_{{n,m+1}}+u_{{n,m-2}} \right) u_{{n,m}} 
            \\
            + 
            \left( -{ \alpha_{3}}+u_{{n,m+1}} \right) u_{{n,m-2}}+u_{{n,m+2}} 
            \left( { \alpha_{3}}-u_{{n,m-1}} \right) 
            \end{gathered}\right]
                }{\left( -u_{{n,m+2}}+u_{{n,m}} \right)  
            \left( -u_{{n,m-2}}+u_{{n,m}} \right)  \left( u_{{n,m-1}}-u_{{n,m+1}} \right)}}
        \end{aligned}
        \label{eq:W2tH2e}
    \end{equation}
    \label{eq:tH2eint}
\end{subequations}

\begin{remark}
    \label{rem:th2e}
     {We note that \tHeq{2} equation possesses
    an autonomous sub-case when $\varepsilon=0$.
    We denote this sub-case by \tHeq[=0]{2}.
    In this sub-case the equations defining the
    first integrals become singular and so
    the first integrals become simpler:
    \begin{subequations}
        \begin{align}
            W_{1}^{\varepsilon=0}&=\left( -1 \right)^{m}
            \frac {2 \alpha_2-u_{n-1,m}+2 u_{n,m}-u_{{n+1,m}}}{u_{{n-1,m}}-u_{{n+1,m}}},
            \label{eq:W1tH2e0}
            \\
            W_{2}^{\varepsilon=0}&=
            {\frac { \left( u_{{n,m+1}}-u_{{n,m-1}} \right)
            \left( u_{{n,m+2}}-u_{{n,m}} \right) }{u_{{n,m}}+u_{{n,m+1}}-\alpha_3}}.
            \label{eq:W2tH2e0}
        \end{align}
        \label{eq:tH2e0int}
    \end{subequations}
    The first integral in the $n$-direction 
    \eqref{eq:W1tH2e0}
    is still non-autonomous, although the equation
    is autonomous, but now it is a three-point, second order
    first integral. 
    On the contrary, the first integral 
    in the $m$-direction \eqref{eq:W2tH2e0}
    is a four-point, third order first integral.

    Finally we note that the \tHeq[=0]{2} equation
    is related
    to the equation (1) from \emph{List 3}
    in \cite{GarifullinYamilov2012}:
\begin{equation}
    \left(\hat{u}_{n+1,m+1}-\hat{u}_{n+1,m}  \right)
    \left(\hat{u}_{n,m}-\hat{u}_{n,m+1}  \right)
    +\hat{u}_{n,m}+\hat{u}_{n+1,m}
    +\hat{u}_{n,m+1}+\hat{u}_{n+1,m+1}=0
    \label{eq:1list3}
\end{equation}
through the transformation:
\begin{equation}
    u_{n,m} = -\alpha_{2}\hat{u}_{m,n}+\frac{1}{4}\alpha_{2}
    +\frac{1}{2}\alpha_{3}.
    \label{eq:transfto1}
\end{equation}
Note that in this formula \eqref{eq:transfto1} the two lattice
variables are \emph{exchanged}.
So it was already known in the literature that the \tHeq[=0]{2} 
equation was Darboux integrable.}
\end{remark}

\subsubsection{\tHeq{3} equation}

Let us consider the \tHeq{3} equation as given by \eqref{eq:tH3e}. It has a
four-point, third order 
first integral in the $n$-direction:
\begin{subequations}
    \begin{align}
        W_{1} & 
        \begin{aligned}[t]
            &=\Fp{m}  \frac { \left( u_{{n-1,m}}-u_{{n+1,m}}
             \right)  \left( -u_{{n+2,m}}+u_{{n,m}} \right) }{{{ \alpha_{2}}}^{4}{
            \epsilon}^{2}{\delta}^{2}-{{ \alpha_{2}}}^{3}u_{{n+1,m}}u_{{n,m}}+ \left( {
            u_{{n,m}}}^{2}+u_{n+1,m}^{2}-2{\epsilon}^{2}{\delta}^{2}
             \right) {{ \alpha_{2}}}^{2}-{ \alpha_{2}}u_{{n,m}}u_{{n+1,m}}+{\epsilon}^{2}{
            \delta}^{2}}
            \\
            &-\Fm{m} {\frac { \left(u_{{n+1,m}} -u_{{n-1,m}} \right)  
            \left( u_{{n+2,m}}-u_{{n,m}} \right) }{{
             \alpha_{2}} \left( -u_{{n-1,m}}+{ \alpha_{2}}u_{{n,m}} \right)  \left( -u_{
            {n+2,m}}+u_{{n+1,m}}{ \alpha_{2}} \right) }}
        \end{aligned}
        \label{eq:W1tH3e}
    \end{align}
    and a five-point, fourth order 
    first integral in the $m$-direction:
    \begin{align}
        W_{2} &= 
        \begin{aligned}[t]
            &\Fp{m}  \alpha \frac { \left( u_{{n,m-1}}-u_{{n,m+1}}\right)^{2} 
            \left(u_{{n,m+2}}- u_{{n,m}} \right)  
            \left(u_{{n,m}}  -u_{{n,m-2}}\right) }{%
                \left({\delta}^{2}\alpha_{3}^{2}+u_{{n,m-1}}^{2}{\epsilon}^{2} 
                -{ \alpha_{3}}u_{{n,m-1}}u_{{n,m}}\right)  
                \left( {\delta}^{2}\alpha_{3}^{2}+u_{{n,m+1}}^{2}{\epsilon}^{2}
                -{ \alpha_{3}}u_{{n,m}}u_{{n,m+1}}
                 \right) } 
            \\
            &-\Fm{m}\beta
            \frac {\left[
                \begin{gathered}
                -u_{n,m}^{2}{ \alpha_{3}}u_{{n,m-1}}+u_{n,m}^{2}{ \alpha_{3}}
            u_{{n,m+1}}-u_{n,m}^{2}{\epsilon}^{2}u_{{n,m+2}}
            \\
            +u_{n,m}^{2}{\epsilon}^{2}u_{{n,m-2}}+{ \alpha_{3}}u_{{n,m}}u_{{n,m-1}}u_{{n,m+2}}
            -{\alpha_{3}}u_{{n,m}}u_{{n,m-2}}u_{{n,m+1}}
            \\
            -{\delta}^{2}{{ \alpha_{3}}}^{2}u_{{n,m+2}}+{\delta}^{2}{{ \alpha_{3}}}^{2}u_{{n,m-2}}
            \end{gathered}\right]}{ \left( u_{{n,m}}-u_{{n,m+2}} \right)  
                \left( -u_{{n,m-2}}+u_{{n,m}} \right)  \left( u_{{n,m-1}}-u_{{n,m+1}} \right)}.
        \end{aligned}
        \label{eq:W2tH3e}
    \end{align}
    \label{eq:tH3eint}
\end{subequations}

\begin{remark}
    With the same notation as in  {Remark \ref{rem:th2e}, 
    we note that also the \tHeq{3} equation has an
    autonomous sub-case of the equation \tHeq{3} if $\varepsilon=0$, 
    namely, the \tHeq[=0]{3} equation.
    In this sub-case the equations defining the
    first integrals become singular and so
    the first integrals become simpler:
    \begin{subequations}
        \begin{align}
            W_{1}^{\varepsilon=0} &=
            \left( -1 \right)^{m} \left[ \frac{\alpha_{2}u_{n,m}-u_{n-1,m}}{%
                u_{n+1,m}-u_{n-1,m}}
            +\frac{1}{2} \right],
            \label{eq:W1tH3e0}
            \\
            W_{2}^{\varepsilon=0} &=
            \frac { \left(u_{{n,m+2}} -u_{{n,m}}\right) u_{{n,m-1}}
            -u_{{n,m+1}}u_{{n,m+2}}+\alpha_{3}\delta^{2}}{%
                \alpha_{3}{\delta}^{2}-u_{{n,m}}u_{{n,m+1}}}.
            \label{eq:W2tH3e0}
        \end{align}
        \label{eq:tH3e0int}
    \end{subequations}
    The first integral in the $n$-direction 
    \eqref{eq:W1tH3e0}
    is still non-autonomous, although the equation
    is autonomous, but now it is a three-point, second order
    first integral. 
    On the contrary, the first integral 
    in the $m$-direction \eqref{eq:W2tH3e0}
    is a four-point, third order first integral.

    Finally we note that the \tHeq[=0]{3} equation
    is related to equation (2) from \emph{List 3}
    in \cite{GarifullinYamilov2012}:
    \begin{equation}
        \hat{u}_{n+1,m+1}\left( \hat{u}_{n,m} +b_{2}\hat{u}_{n,m+1}\right)
        +\hat{u}_{n+1,m}\left( b_{2}\hat{u}_{n,m}+\hat{u}_{n,m+1} \right)+c_{4}=0
        \label{eq:2list3}
    \end{equation}
    through the inversion of two lattice parameters
    $u_{n,m}=\hat{u}_{m,n}$ and the choice of parameters:
    \begin{equation}
        b_{2} = -\frac{1}{\alpha_{2}},
        \quad
        c_{4} = \frac{\delta^{2}\alpha_{3}\left( 1-\alpha_{2}^{2} \right)}{\alpha_{2}}.
        \label{eq:gyparam}
    \end{equation}
    So it was already known in the literature that the \tHeq[=0]{3} 
    equation was Darboux integrable.
} 
    
%
\end{remark}

\begin{remark}
    As a final remark we can say that the first integrals
    of the \tHeq{2}  and \tHeq{3} 
    equations have the same order in each direction. Furthermore,
    they share the important property that in the direction
    $m$, which is the direction of the non-autonomous factors
    $F_{m}^{(\pm)}$, the $W_{2}$ integrals are built up from
    two different ``sub''-integrals as in the known case
    of the \tHeq{1} equation.
\end{remark}

\subsection{\Hsechs~equations}

In this Subsection we consider the equations of the family
\Hsechs~introduced in \cite{Boll2011,Boll2012a,Boll2012b}. 
We present
their non-autonomous form, as it is given in \cite{GSL_general}:
\begin{subequations}
    \begin{align}
        _{1}D_{2} &\colon
        \begin{aligned}[t]
        &\phantom{+}\left( F_{n+m}^{\left(-\right)}-\delta_{1} F_{n}^{\left(+\right)} F_{m}^{\left(-\right)}+\delta_{2} F_{n}^{\left(+\right)} F_{m}^{\left(+\right)}\right)u_{n,m}
        \\
        &+\left( F_{n+m}^{\left(+\right)}-\delta_{1} F_{n}^{\left(-\right)} F_{m}^{\left(-\right)}+\delta_{2} F_{n}^{\left(-\right)} F_{m}^{\left(+\right)}\right)u_{n+1,m}
        \\ 
        &+\left( F_{n+m}^{\left(+\right)}-\delta_{1} F_{n}^{\left(+\right)} F_{m}^{\left(+\right)}+\delta_{2} F_{n}^{\left(+\right)} F_{m}^{\left(-\right)}\right)u_{n,m+1}
        \\
        &+\left( F_{n+m}^{\left(-\right)}-\delta_{1} F_{n}^{\left(-\right)} F_{m}^{\left(+\right)}+\delta_{2} F_{n}^{\left(-\right)} F_{m}^{\left(-\right)}\right)u_{n+1,m+1}
        \\ 
        &+\delta_{1}\left( F_{m}^{\left(-\right)}u_{n,m}u_{n+1,m}+ F_{m}^{\left(+\right)}u_{n,m+1}u_{n+1,m+1}\right)
        \\
        &+ F_{n+m}^{\left(+\right)}u_{n,m}u_{n+1,m+1}
        + F_{n+m}^{\left(-\right)}u_{n+1,m}u_{n,m+1}=0,
        \end{aligned}
        \label{eq:1D2}
        \\
        _{2}D_{2} &\colon
        \begin{aligned}[t]
            &\phantom{+}\left(F_{m}^{\left(-\right)}-\delta_{1}F_{n}^{\left(+\right)}F_{m}^{\left(-\right)}+\delta_{2}F_{n}^{\left(+\right)}F_{m}^{\left(+\right)}-\delta_{1} \lambda F_{n}^{\left(-\right)}F_{m}^{\left(+\right)}\right)u_{n,m}
        \\
        &+\left(F_{m}^{\left(-\right)}-\delta_{1}F_{n}^{\left(-\right)}F_{m}^{\left(-\right)}+\delta_{2}F_{n}^{\left(-\right)}F_{m}^{\left(+\right)}-\delta_{1} \lambda F_{n}^{\left(+\right)}F_{m}^{\left(+\right)}\right)u_{n+1,m}
        \\
        &+\left(F_{m}^{\left(+\right)}-\delta_{1}F_{n}^{\left(+\right)}F_{m}^{\left(+\right)}+\delta_{2}F_{n}^{\left(+\right)}F_{m}^{\left(-\right)}-\delta_{1} \lambda F_{n}^{\left(-\right)}F_{m}^{\left(-\right)}\right)u_{n,m+1}
        \\
        &+\left(F_{m}^{\left(+\right)}-\delta_{1}F_{n}^{\left(-\right)}F_{m}^{\left(+\right)}+\delta_{2}F_{n}^{\left(-\right)}F_{m}^{\left(-\right)}-\delta_{1} \lambda F_{n}^{\left(+\right)}F_{m}^{\left(-\right)}\right)u_{n+1,m+1}
        \\
        &+\delta_{1}\left(F_{n+m}^{\left(-\right)}u_{n,m}u_{n+1,m+1}+F_{n+m}^{\left(+\right)}u_{n+1,m}u_{n,m+1}\right)
        \\ 
        &+F_{m}^{\left(+\right)}u_{n,m}u_{n+1,m}+F_{m}^{\left(-\right)}u_{n,m+1}u_{n+1,m+1}
        -\delta_{1}\delta_{2}\lambda=0,
        \end{aligned}
        \label{eq:2D2}
        \\
        _{3}D_{2} &\colon
        \begin{aligned}[t]
            &\phantom{+}\left(F_{m}^{\left(-\right)}-\delta_{1}F_{n}^{\left(-\right)}F_{m}^{\left(-\right)}+\delta_{2}F_{n}^{\left(+\right)}F_{m}^{\left(+\right)}-\delta_{1} \lambda F_{n}^{\left(-\right)}F_{m}^{\left(+\right)}\right)u_{n,m}
        \\
        &+\left(F_{m}^{\left(-\right)}-\delta_{1}F_{n}^{\left(+\right)}F_{m}^{\left(-\right)}+\delta_{2}F_{n}^{\left(-\right)}F_{m}^{\left(+\right)}-\delta_{1} \lambda F_{n}^{\left(+\right)}F_{m}^{\left(+\right)}\right)u_{n+1,m}
        \\
        &+\left(F_{m}^{\left(+\right)}-\delta_{1}F_{n}^{\left(-\right)}F_{m}^{\left(+\right)}+\delta_{2}F_{n}^{\left(+\right)}F_{m}^{\left(-\right)}-\delta_{1} \lambda F_{n}^{\left(-\right)}F_{m}^{\left(-\right)}\right)u_{n,m+1}
        \\
        &+\left(F_{m}^{\left(+\right)}-\delta_{1}F_{n}^{\left(+\right)}F_{m}^{\left(+\right)}+\delta_{2}F_{n}^{\left(-\right)}F_{m}^{\left(-\right)}-\delta_{1} \lambda F_{n}^{\left(+\right)}F_{m}^{\left(-\right)}\right)u_{n+1,m+1}
        \\
        &+\delta_{1}\left(F_{n}^{\left(-\right)}u_{n,m}u_{n,m+1}+F_{n}^{\left(+\right)}u_{n+1,m}u_{n+1,m+1}\right) 
        \\ 
        &+F_{m}^{\left(-\right)}u_{n,m+1}u_{n+1,m+1}
        +F_{m}^{\left(+\right)}u_{n,m}u_{n+1,m}-\delta_{1}\delta_{2}\lambda=0,
        \end{aligned}
        \label{eq:3D2}
        \\
        D_{3} &\colon
        \begin{aligned}[t]
            &\phantom{+}F_{n}^{\left(+\right)}F_{m}^{\left(+\right)}u_{n,m}+F_{n}^{\left(-\right)}F_{m}^{\left(+\right)}u_{n+1,m}
            +F_{n}^{\left(+\right)}F_{m}^{\left(-\right)}u_{n,m+1}
            \\
            &+F_{n}^{\left(-\right)}F_{m}^{\left(-\right)}u_{n+1,m+1}
            +F_{m}^{\left(-\right)}u_{n,m}u_{n+1,m}
            \\
            &+F_{n}^{\left(-\right)}u_{n,m}u_{n,m+1}+F_{n+m}^{\left(-\right)}u_{n,m}u_{n+1,m+1}
        \\
        &+F_{n+m}^{\left(+\right)}u_{n+1,m}u_{n,m+1}+F_{n}^{\left(+\right)}u_{n+1,m}u_{n+1,m+1}
        \\
        &+F_{m}^{\left(+\right)}u_{n,m+1}u_{n+1,m+1}=0,
        \end{aligned}
        \label{eq:D3}
        \\
        _{1}D_{4} &\colon
        \begin{aligned}[t]
            &\phantom{+}\delta_{1}\left(F_{n}^{\left(-\right)}u_{n,m}u_{n,m+1}+F_{n}^{\left(+\right)}u_{n+1,m}u_{n+1,m+1}\right)\\
            &+\delta_{2}\left(F_{m}^{\left(-\right)}u_{n,m}u_{n+1,m}+F_{m}^{\left(+\right)}u_{n,m+1}u_{n+1,m+1}\right)\\
            &+u_{n,m}u_{n+1,m+1}+u_{n+1,m}u_{n,m+1}+\delta_{3}=0,
        \end{aligned}
        \label{eq:1D4}
        \\
        _{2}D_{4} &\colon
        \begin{aligned}[t]
            &\phantom{+}\delta_{1}\left(F_{n}^{\left(-\right)}u_{n,m}u_{n,m+1}
            +F_{n}^{\left(+\right)}u_{n+1,m}u_{n+1,m+1}\right)
            \\
            &+\delta_{2}\left(F_{n+m}^{\left(-\right)}u_{n,m}u_{n+1,m+1}
            +F_{n+m}^{\left(+\right)}u_{n+1,m}u_{n,m+1}\right)
            \\
            &+u_{n,m}u_{n+1,m}+u_{n,m+1}u_{n+1,m+1}+\delta_{3}=0,
        \end{aligned}
        \label{eq:2D4}
    \end{align}
    \label{eq:h6}
\end{subequations}
where the functions $F_{k}^{(\pm)}$ are given by 
\eqref{eq:fk}.

We now give formulae for the first integrals
of these equations.

\subsubsection{$_{1}D_{2}$ equation}

In case of the $_{1}D_{2}$ equation given by formula \eqref{eq:1D2}, 
we have the following  three-point, second order first integrals:
\begin{subequations}
    \begin{align}
        W_{1} &
        \begin{aligned}[t]
            &={ \Fppp}{ \alpha} {\frac { \left[  \left( 1+{ \delta_{2}}\right) u_{{n,m}}+u_{{n+1,m}} \right] { 
            \delta_{1}}-u_{{n,m}}}{\left[  \left( 1+{ \delta_{2}} \right) 
            u_{{n,m}}+u_{{n-1,m}} \right] { \delta_{1}}-u_{{n,m}} }}
            \\
            &+{ \Fpmm}{ \alpha} {\frac {1+ \left( u_{{n+1,m}}-1 \right) { \delta_{1}}}{ 
                1+ \left( u_{{n-1,m}}-1 \right) { \delta_{1}}}}
            \\
            &+{ \Fmpm}{ \beta} \left( u_{{n+1,m}} -u_{{n-1,m}}\right) 
            \\
            &-{ \Fmmp}{ \beta}{\frac { \left( u_{{n+1,m}} -u_{{n-1,m}}\right)  
            \left[ 1- \left( 1-u_{{n,m}} \right) { \delta_{1}} \right] }{{ \delta_{2}}+u_{{n,m}}}},
        \end{aligned}
        \label{eq:W11D2}
        \\
        W_{2} &
        \begin{aligned}[t]
            &= { \Fppp}{ \alpha} {\frac {u_{{n,m+1}} -u_{{n,m-1}}}{
                u_{{n,m}}+{ \delta_{1}}u_{{n,m-1}}}}
            \\
            &+{ \Fpmm}{ \beta} \left( u_{{n,m+1}} -u_{{n,m-1}}\right) 
            \\
            &-{ \Fmpm}{ \alpha} \frac {u_{{n,m+1}}-u_{{n,m-1}}}{
                1+\delta_{1}\left(u_{n,m+1} -1\right)}
            \\
            &-{ \Fmmp}{ \beta}{\frac {u_{{n,m+1}} -u_{{n,m-1}}}{%
                \delta_{2}+u_{{n,m}}}}.
        \end{aligned}
        \label{eq:W21D2}
    \end{align}
    \label{eq:1D2int}
\end{subequations}

 {
\begin{remark}
    \label{rem:1D2}
    We remark that, when $\delta_{1} \to0$, the
    first integral \eqref{eq:W11D2} is
    singular, since the coefficient at $\alpha$
    approaches a constant. In this particular
    case, it can be shown that the first integrals
    are given by:
    \begin{subequations}
        \begin{align}
            W_{1}^{(0,\delta_{2})} &
            \begin{aligned}[t]
                &=\Fppp \alpha
                \frac {u_{n+1,m}-u_{n-1,m}}{u_{n,m}}
                \\
                &-\Fpmm \alpha\left( u_{{n+1,m}}-u_{{n-1,m}}\right) 
                \\
                &+\Fmpm \beta \left( u_{{n+1,m}}-u_{{n-1,m}} \right) 
                \\
                &+ \Fmmp\beta{\frac {u_{{n-1,m}}-u_{{n+1,m}}}{{  \delta_2}+u_{{n,m}}}},
            \end{aligned}
            \label{eq:W11D20d2}
            \\
        W_{2}^{(0,\delta_{2})} &
        \begin{aligned}[t]
            &= { \Fppp}{ \alpha} {\frac {u_{{n,m+1}} -u_{{n,m-1}}}{
                u_{{n,m}}}}
                \\
            &+{ \Fpmm}{ \beta} \left( u_{{n,m+1}} -u_{{n,m-1}}\right) 
            \\
            &-{ \Fmpm}{ \alpha} \left(u_{n,m+1}-u_{n,m-1}\right)
            \\
            &-{ \Fmmp}{ \beta}{\frac {u_{{n,m+1}} -u_{{n,m-1}}}{%
                \delta_{2}+u_{{n,m}}}}.
        \end{aligned}
            \label{eq:W21D20d2}
        \end{align}
        \label{eq:1D20d2int}
    \end{subequations}
    Note that, as the limit in \eqref{eq:W21D2}
    is not singular, then \eqref{eq:W21D20d2} can be
    obtained directly from \eqref{eq:W21D2}.

    If $\delta_{1} \to\left( 1+\delta_{2} \right)^{-1}$, then  
    the limit of the first integral \eqref{eq:W21D2} is
    not singular. However, it can be seen
    from equations defining the first
    integral in the $m$-direction that,
    in this case, there might exist a
    two-point, first order first integral.
    Carrying out the computations, we obtain    
    that if $\delta_{1}=\left( 1+\delta_{2} \right)^{-1}$, 
    the first integrals
    are given by:
    \begin{subequations}
        \begin{align}
            W_{1}^{(\left( 1+\delta_{2} \right)^{-1},\delta_{2})} &
            \begin{aligned}[t]
                &=\Fppp\alpha{\frac { u_{{n+1,m}}}{u_{{n-1,m}}}}
                +\Fpmm\alpha{\frac { {\delta_{2}}+u_{{n+1,m}} }{\delta_{2}+u_{n-1,m}}}
                \\
                &+ \Fmpm \beta\left( u_{{n+1,m}}-u_{{n-1,m}} \right)
                \\
                &-\Fmmp\beta{\frac{\left( u_{{n+1,m}}-u_{{n-1,m}} \right)}{{\delta_{2}}+1}},
            \end{aligned}
            \label{eq:W11D2d1d2}
            \\
        W_{2}^{\left(\left( 1+\delta_{2} \right)^{-1},\delta_{2}\right)} &
        \begin{aligned}[t]
            &=\Fppp \alpha  \left[ \left(1+{  \delta_{2}}\right) u_{{n,m}}+u_{{n,m+1}} \right]
            \\
            &+{  \Fpmm} \beta {\frac {  u_{{n,m}}+\left(1+{  \delta_{2}}\right) u_{{n,m+1}}  }{{  \delta_{2}}+1}}
            \\
            &-{  \Fmpm} \alpha
            {\frac {  \left( {  \delta_{2}}+1 \right) u_{{n,m}}}{{  \delta_{2}}+u_{{n,m+1}}}}
            -{  \Fmmp} \beta{\frac { u_{{n,m+1}}}{{  \delta_{2}}+u_{{n,m}}}}.            
        \end{aligned}
            \label{eq:W21D2d1d2}
        \end{align}
        \label{eq:1D2d1d2int}
    \end{subequations}
    So, unlike the case 
    $\delta_{1}\neq\left( 1+\delta_{2} \right)^{-1}$, 
    the first integral in the $m$-direction 
    \eqref{eq:W21D2d1d2} is a two-point, first
    order first integral. 
    On the contrary,
    the first integral in the $n$-direction
    \eqref{eq:W11D2d1d2} is still a three-point,
    second order first integral, which can
    be obtained from the complete form
    \eqref{eq:W11D2} by substituting 
    $\delta_{1}=\left( 1+\delta_{2} \right)^{-1}$.
\end{remark}
}

\subsubsection{$_{2}D_{2}$ equation}

In the case of the $_{2}D_{2}$ equation given by \eqref{eq:2D2}, 
we have the following three-point, second order first integrals:
\begin{subequations}
    \begin{align}
        W_{1} &
        \begin{aligned}[t]
            &={ \Fppp}{\alpha} {\frac {{ \delta_{2}}+u_{{n+1,m}}}{{ \delta_{2}}+u_{{n-1,m}}}}
            \\
            &+{ \Fpmm}{\alpha} {\frac { \left( 1- \left( 1+{ \delta_{2}} \right) { 
                \delta_{1}} \right) u_{{n,m}}+u_{{n+1,m}}}{ \left( 1
                - \left( 1+{ \delta_{2}} \right) { \delta_{1}}\right) u_{{n,m}}+u_{{n-1,m}}}} 
            \\
            &+{ \Fmpm}{\beta} {\frac { \left(u_{{n+1,m}} -u_{{n-1,m}} \right)  
            \left( u_{{n,m}}+{ \delta_{2}} \right) }{1+ \left( -1+u_{{n,m}} \right) { \delta_{1}}}}
            \\
            &-{ \Fmmp}{\beta} \left( u_{{n+1,m}} -u_{{n-1,m}}\right),
        \end{aligned}
        \label{eq:W12D2}
        \\
        W_{2} &
        \begin{aligned}[t]
            &={ \Fppp}{\alpha} \left( u_{{n,m+1}} -u_{{n,m-1}}\right) 
            \\
            &-{ \Fpmm}{\beta} {\frac {u_{{n,m+1}}-u_{{n,m-1}}}{ 
                \left( \lambda-u_{{n,m}} \right) { \delta_{1}}-u_{{n,m-1}}}}
            \\
            &-{\Fmpm}{\alpha} {\frac {u_{{n,m+1}} -u_{{n,m-1}}}{
                1+ \left( -1+u_{{n,m}} \right) { \delta_{1}}}}
            \\
            &-{ \Fmmp}{\beta}{\frac {u_{{n,m+1}}-u_{{n,m-1}}}{u_{{n,m+1}}+{ \delta_{2}}}}.
        \end{aligned}
        \label{eq:W22D2}
    \end{align}
    \label{eq:2D2int}
\end{subequations}

 {
\begin{remark}
    \label{rem:2D2}
    We remark that if $\delta_{1} \to0$, 
    the first integral \eqref{eq:W12D2} is
    not singular. However it can be seen
    from equations defining the first
    integral in the $n$-direction that
    in this case there might exist a
    two-point, first order first integral.
    Carrying out the computations, we obtain    
    that if $\delta_{1}=0$, the first integrals
    are given by:
    \begin{subequations}
        \begin{align}
            W_{1}^{(0,\delta_{2})} &
            \begin{aligned}[t]
                &= \Fppp \alpha 
                \left(  {  \delta_2}+u_{{n+1,m}} \right) u_{{n,m}}
                -\Fpmm \alpha \left( u_{{n+1,m}}+u_{{n,m}} \right) 
                \\
                &+ \Fmpm \beta \left(\delta_2+ u_{{n,m}} \right)u_{{n+1,m}}
                -\Fmmp \beta \left( u_{{n+1,m}}+u_{{n,m}} \right) ,
            \end{aligned}
            \label{eq:W12D20d2}
            \\
        W_{2}^{(0,\delta_{2})} &
        \begin{aligned}[t]
            &={ \Fppp}{\alpha} \left( u_{{n,m+1}} -u_{{n,m-1}}\right)
            +{ \Fpmm}{\beta} {\frac {u_{{n,m+1}}}{u_{{n,m-1}}}}
            \\
            &-{\Fmpm}{\alpha} \left( {u_{{n,m+1}} -u_{{n,m-1}}}\right)
            \\
            &-{ \Fmmp}{\beta}{\frac {u_{{n,m+1}}-u_{{n,m-1}}}{u_{{n,m+1}}+{ \delta_{2}}}} .
        \end{aligned}
            \label{eq:W22D20d2}
        \end{align}
        \label{eq:2D20d2int}
    \end{subequations}
    So, differently from the case 
    $\delta_{1}\neq0$,
    the first integral in the $n$-direction 
    \eqref{eq:W12D20d2} is a two-point, first
    order first integral. 
    On the contrary,
    the first integral in the $m$-direction
    \eqref{eq:W22D20d2} is still a three-point,
    second order first integral, which can
    be obtained from the complete form
    \eqref{eq:W22D2} by substituting 
    $\delta_{1}=0$.

    Moreover we note that
    if $\delta_{1} \to\left( 1+\delta_{2} \right)^{-1}$, 
    the limit of the first integral \eqref{eq:W22D2} is
    not singular. However, it can be seen
    from equations defining the first
    integral in the $m$-direction that
    in this case there might exist a
    two-point, first order first integral.
    Carrying out the computations, we obtain    
    that if $\delta_{1}=\left( 1+\delta_{2} \right)^{-1}$, 
    the first integrals
    are given by:
    \begin{subequations}
        \begin{align}
            W_{1}^{\left(\left( 1+\delta_{2} \right)^{-1},\delta_{2}\right)} &
            \begin{aligned}[t]
                &={\Fppp} {  \alpha}
                {\frac {{  \delta_{2}}+u_{{n+1,m}}}{{\delta_{2}}+u_{{n-1,m}}}} 
                \\
                &+\Fpmm{  \alpha} {\frac {u_{{n+1,m}}}{u_{{n-1,m}}}}
                \\
                &-\Fmpm{\beta} \left( u_{{n-1,m}}-u_{{n+1,m}} \right)
                \\
                &-{\Fmmp} {\beta} \frac{u_{{n+1,m}}-u_{{n-1,m}}}{{  1+\delta_{2}}},
            \end{aligned}
            \label{eq:W12D2d1d2}
            \\
        W_{2}^{\left(\left( 1+\delta_{2} \right)^{-1},\delta_{2}\right)} &
        \begin{aligned}[t]
            &={\Fppp} \alpha  \left[ \left(1+{  \delta_{2}}\right) u_{{n,m}}+u_{{n,m+1}} \right]
            \\
            &+\Fpmm\beta\left[ u_{{n,m}}+\left(1+{\delta_{2}}\right) u_{{n,m+1}} \right]
            \\
            &+{\Fmpm}\alpha{\frac {{  \delta_{2}} \lambda-\left(1+{\delta_{2}}\right) u_{{n,m+1}}+\lambda u_{{n,m}}}{%
                u_{{n,m}}+{  \delta_{2}}}}
            \\
            &+\Fmmp\beta{\frac{{  \delta_{2}} \lambda -\left(1+{  \delta_{2}}\right) u_{{n,m}}+\lambda u_{{n,m+1}}}{
                u_{{n,m+1}}+{\delta_{2}}}}.
        \end{aligned}
            \label{eq:W22D2d1d2}
        \end{align}
        \label{eq:2D2d1d2int}
    \end{subequations}
    So, differently from the case 
    $\delta_{1}\neq\left( 1+\delta_{2} \right)^{-1}$,
    the first integral in the $m$-direction 
    \eqref{eq:W22D2d1d2} is a two-point, first
    order first integral. 
    On the contrary,
    the first integral in the $n$-direction
    \eqref{eq:W12D2d1d2} is still a three-point,
    second order first integral, which can
    be obtained from the complete form
    \eqref{eq:W12D2} by substituting 
    $\delta_{1}=\left( 1+\delta_{2} \right)^{-1}$.
\end{remark}
}

\subsubsection{$_{3}D_{2}$ equation}

In the case of the $_{3}D_{2}$ equation given by \eqref{eq:3D2}, 
we have the following three-point, second order first integrals:
\begin{subequations}
    \begin{align}
        W_{1} &
        \begin{aligned}[t]
            &={ \Fppp}{\alpha } {\frac { \left( u_{{n-1,m}}+{ \delta_{2}} \right)  
            \left[ 1+ \left( u_{{n+1,m}}-1 \right) { \delta_{1}} \right] }{ 
                \left( u_{{n+1,m}}+{ \delta_{2}} \right)  
               \left[ 1+ \left( u_{{n-1,m}}-1 \right) { \delta_{1}} \right] }}
            \\
            &+{\Fpmm}{ \alpha}\frac {u_{{n,m}}+\left( 1-\delta_{1}-\delta_{1}\delta_{2} \right)u_{{n-1,m}}}{%
            u_{{n,m}}+\left( 1-\delta_{1}-\delta_{1}\delta_{2} \right)u_{{n+1,m}}}
            \\
            &+{\Fmpm}{ \beta} \left(u_{{n+1,m}}- u_{{n-1,m}}\right) 
            \left( { \delta_{2}}+u_{{n,m}} \right) 
            \\
            &-{ \Fmmp}{ \beta}\left(u_{{n+1,m}} -u_{{n-1,m}}\right),
        \end{aligned}
        \label{eq:W13D2}
        \\
        W_{2} &
        \begin{aligned}[t]
            &={ \Fppp}{ \alpha} \left( u_{{n,m+1}} -u_{{n,m-1}}\right) 
            \\
            &-{\Fpmm}{ \beta} {\frac {u_{{n,m+1}}- u_{{n,m-1}}}{
                \lambda \left( 1+{ \delta_{2}} \right) 
                {\delta_{1}}^{2}- \left[  \left( 1+{ \delta_{2}} \right) u_{{n,m-1}}+u_{{n,m}}+\lambda \right]
                {\delta_{1}}+u_{{n,m-1}}}}
            \\
            &+{ \Fmpm}{ \alpha}  \left( u_{{n,m-1}}-u_{{n,m+1}} \right)  
            \left[ 1+ \left( u_{{n,m}}-1 \right) { \delta_{1}} \right]
            \\
            &+{ \Fmmp}{ \beta} {\frac {u_{{n,m+1}} - u_{{n,m-1}}}{ 
                \left( { \delta_{2}}+u_{{n,m+1}} \right)  
                \left[ 1+\left(1-{ \delta_{1}}\right)u_{{n,m-1}} \right] }}.
        \end{aligned}
        \label{eq:W23D2}
    \end{align}
    \label{eq:3D2int}
\end{subequations}

 {
    \begin{remark}
    We remark that if $\delta_{1} \to0$, 
    the first integral \eqref{eq:W13D2} is
    not singular. However, it can be seen
    from equations defining the first
    integral in the $n$-direction that
    in this case there might exist a
    two-point, first order first integral.
    Carrying out the computations, we obtain    
    that if $\delta_{1}=0$, the first integrals
    are given by:
    \begin{subequations}
        \begin{align}
            W_{1}^{(0,\delta_{2})} &
            \begin{aligned}[t]
                &=\Fppp \alpha u_{{n,m}} \left( {\delta_{2}}+u_{{n+1,m}} \right) 
                -\Fpmm \alpha  \left( u_{{n+1,m}}+u_{{n,m}} \right)
                \\
                &+\Fmpm \beta u_{{n+1,m}} \left({  \delta_{2}}+ u_{{n,m}} \right) 
                -\Fmmp \beta  \left( u_{{n+1,m}}+u_{{n,m}} \right) ,               
            \end{aligned}
            \label{eq:W13D20d2}
            \\
        W_{2}^{(0,\delta_{2})} &
        \begin{aligned}[t]
            &=\Fppp \alpha  \left( u_{{n,m+1}}-u_{{n,m-1}} \right) 
            +\Fpmm \beta{\frac { u_{{n,m+1}}}{u_{{n,m-1}}}}
            \\
            &-\Fmpm\alpha  \left( u_{{n,m+1}}-u_{{n,m-1}} \right) 
            +\Fmmp \beta{\frac {{  \delta_{2}}+  u_{{n,m-1}} }{\delta_{2}+u_{{n,m+1}}}}.
        \end{aligned}
            \label{eq:W23D20d2}
        \end{align}
        \label{eq:2D30d2int}
    \end{subequations}
    So, differently from the case 
    $\delta_{1}\neq0$,
    the first integral in the $n$-direction 
    \eqref{eq:W13D20d2} is a two-point, first
    order first integral. 
    On the contrary,
    the first integral in the $m$-direction
    \eqref{eq:W23D20d2} is still a three-point,
    second order first integral, which can
    be obtained from the complete form
    \eqref{eq:W23D2} by substituting 
    $\delta_{1}=0$.

        Moreover we remark that if 
        $\delta_{1} \to \left( 1+\delta_{2} \right)^{-1}$,
        the first integral \eqref{eq:W13D2}
        is singular, since the coefficient at $\alpha$
        approaches a constant.
        In this particular case the first integrals
        are given by:
        \begin{subequations}
            \begin{align}
                W_{1}^{\left(\left( 1+\delta_{2} \right)^{-1},\delta_{2}\right)} &
                \begin{aligned}[t]
                   &= {\Fppp} {  \alpha} {\frac {u_{{n+1,m}}-u_{{n-1,m}}}{ 
                        \left( {\delta_2}+u_{{n+1,m}} \right)  \left( {  \delta_{2}}+u_{{n-1,m}} \right) }}
                        \\
                  &+{\Fpmm} {  \alpha} {\frac {u_{{n+1,m}}-u_{{n-1,m}}}{ \left( {  \delta_{2}}+1 \right) u_{{n,m}}}}
                  \\
                  &-{\Fmpm} {  \beta} \left( u_{{n-1,m}}-u_{{n+1,m}} \right)  
                  \left( {  \delta_{2}}+u_{{n,m}} \right)
                  \\
                  &-{  \Fmmp} {  \beta} \left( u_{{n+1,m}}-u_{{n-1,m}} \right), 
                \end{aligned}
                \label{eq:W13D2d1d2}
                \\
                W_{2}^{\left(\left( 1+\delta_{2} \right)^{-1},\delta_{2}\right)} &
                \begin{aligned}[t]
                    &={\Fppp} {  \alpha} \left( u_{{n,m+1}}-u_{{n,m-1}} \right) 
                    \\
                    &+{\Fpmm} {  \beta} {\frac {u_{{n,m+1}}-u_{{n,m-1}}}{u_{{n,m}}}}
                    \\
                    &-{\Fmpm} {  \alpha} {\frac { \left( u_{{n,m+1}}-u_{{n,m-1}} \right)  
                    \left( {  \delta_{2}}+u_{{n,m}} \right) }{{  \delta_{2}}+1}}
                    \\
                    &+{\Fmmp} {  \beta}{\frac {u_{{n,m+1}}-u_{{n,m-1}}}{ %
                        \left( {  \delta_{2}}+u_{{n,m+1}} \right)  \left( u_{{n,m-1}}+{\delta_{2}} \right) }}.
                \end{aligned}
                \label{eq:W23D2d1d2}
            \end{align}
            \label{eq:3D2intd1d2}
        \end{subequations}
    We point out that 
    the first integral in the
    $m$-direction \eqref{eq:W23D2d1d2} 
    can be obtained from the complete form
    \eqref{eq:W23D2} in the limit $\delta_{1}\to \left( 1+\delta_{2} \right)^{-1}$.
    \end{remark}
}

\subsubsection{$D_{3}$ equation}

In the case of the $D_{3}$ equation given by \eqref{eq:D3}, 
we have the following four-point, third order
first integrals:
\begin{subequations}
    \begin{align}
        W_{1} &
        \begin{aligned}[t]
            &={ \Fppp}{ \alpha}  {\frac { \left( u_{{n+1,m}}-u_{{n-1,m}} \right)  
            \left( u_{{n+2,m}}-u_{{n,m}} \right) }{u_{n+1,m}^{2}-u_{{n,m}}}}
            \\
            &+{ \Fpmm}{ \alpha} {\frac { \left( u_{{n+1,m}}-u_{{n-1,m}} \right)  
            \left( u_{{n+2,m}-u_{{n,m}}} \right) }{u_{{n,m}}+u_{{n-1,m}}}}
            \\
            &-{ \Fmpm}{ \beta} \frac { \left( u_{{n+1,m}}-u_{{n-1,m}} \right)  
            \left(u_{{n+2,m}} -u_{{n,m}} \right) }{u_{{n+1,m}}- u_{n,m}^{2}} 
            \\
            &+{ \Fmmp}{ \beta}\frac { \left( u_{{n+1,m}}-u_{{n-1,m}} \right)  
            \left( u_{{n+2,m}}-u_{{n,m}} \right) }{u_{{n+1,m}}+u_{{n+2,m}}},
        \end{aligned}
        \label{eq:W1D3}
        \\
        W_{2} &
        \begin{aligned}[t]
            &={ \Fppp}{ \alpha}  {\frac { \left( u_{{n,m+1}}-u_{{n,m-1}} \right)  
            \left( u_{{n,m+2}}-u_{{n,m}} \right) }{u_{n,m+1}^{2}-u_{{n,m}}}}
            \\
            &-{ \Fpmm}{ \beta} {\frac { \left( u_{{n,m+1}}-u_{{n,m-1}} \right)  
            \left(u_{{n,m+2}} -u_{{n,m}} \right) }{u_{{n,m+1}}-u_{n,m}^{2}}}
            \\
            &+{ \Fmpm}{ \alpha} \frac { \left( u_{{n,m+1}}-u_{{n,m-1}} \right)  
            \left(u_{{n,m+2}} -u_{{n,m}} \right) }{u_{{n,m}}+u_{{n,m-1}}}
            \\
            &+{ \Fmmp}{ \beta}\frac { \left( u_{{n,m+1}}-u_{{n,m-1}} \right)  
            \left( u_{{n,m+2}}-u_{{n,m}} \right) }{u_{{n,m+1}}+u_{{n,m+2}}}.
        \end{aligned}
        \label{eq:W2D3}
    \end{align}
    \label{eq:D3int}
\end{subequations}

\begin{remark}
    The equation $D_{3}$ is invariant under the exchange
    of lattice variables $n\leftrightarrow m$.
    Therefore its $W_{2}$ first integral \eqref{eq:W2D3} can be obtained
    from the $W_{1}$ one \eqref{eq:W1D3} simply by 
    exchanging the indices $n$ and $m$.
\end{remark}

\subsubsection{$_{1}D_{4}$ equation}

In the case of the $_{1}D_{4}$ equation given by \eqref{eq:1D4}, 
we have the following four-point, third order
first integrals:
\begin{subequations}
    \begin{align}
        W_{1} &
        \begin{aligned}[t]
            &={ \Fppp}{\alpha } \frac {u_{n+1,m}^{2}{ \delta_{1}}+u_{{n+1,m}}u_{{n+2,m}}+u_{{n-1,m}} 
            \left( u_{{n,m}}-u_{{n+2,m}} \right) -{ \delta_{2}}{ \delta_{3}}}{%
                u_{n+1,m}\left( \delta_{1}+u_{n,m}\right) -\delta_{2}\delta_{3}}
            \\    
            &+{ \Fpmm}{\alpha} \frac { \left( u_{{n,m}}-u_{{n+2,m}}+{\delta_{1}}u_{{n+1,m}} \right) 
                u_{{n-1,m}}+u_{{n+1,m}}u_{{n+2,m}}}{ 
                    \left( u_{{n,m}}+{ \delta_{1}}u_{{n-1,m}} \right) u_{{n+1,m}}}
            \\
            &+{ \Fmpm}{\beta} \frac { \left(u_{{n+1,m}} -u_{{n-1,m}} \right)  
            \left(u_{{n+2,m}}- u_{{n,m}} \right) }{%
                u_{n,m}^{2}{ \delta_{1}}+u_{{n+1,m}}u_{{n,m}} -{ \delta_{2}}{ \delta_{3}}}
            \\
            &+{ \Fmmp}{ \beta} {\frac { \left( u_{{n+1,m}}-u_{{n-1,m}} \right)  
            \left( u_{{n+2,m}}-u_{{n,m}} \right) }{
                u_{{n,m}} \left( u_{{n+2,m}}{ \delta_{1}}+u_{{n+1,m}} \right) }},
        \end{aligned}
        \label{eq:W11D4}
        \\
        W_{2} &
        \begin{aligned}[t]
            &= { \Fppp}\alpha {\frac {\left(u_{{n,m+2}} -u_{{n,m}}\right) u_{{n,m-1}}+\delta_{1} \delta_{3}
                - \delta_{2}u_{n,m+1}^{2}-u_{{n,m+1}}u_{{n,m+2}}}{
                { \delta_{1}}{ \delta_{3}}-u_{{n,m}}u_{{n,m+1}} -{ \delta_{2}}u_{n,m+1}^{2}}}
            \\
            &-{ \Fpmm}\beta {\frac { \left( u_{{n,m+1}}-u_{{n,m-1}} \right) 
            \left( u_{{n,m+2}}-u_{{n,m}} \right) }{%
                \delta_{1}\delta_{3}-{ \delta_{2}}{u_{{n,m}}}^{2}-u_{{n,m}}u_{{n,m+1}}}}
            \\
            &+{ \Fmpm}\alpha{\frac { \left( u_{{n,m}}-u_{{n,m+2}}+{ \delta_{2}}u_{{n,m+1}} \right) u_{{n,m-1}}
            +u_{{n,m+1}}u_{{n,m+2}}}{ \left( u_{{n,m}}+{ \delta_{2}}u_{{n,m-1}} \right) u_{{n,m+1}}}}
            \\
            &+{ \Fmmp}\beta {\frac { \left(u_{{n,m+1}}- u_{{n,m-1}} \right)  
            \left( u_{{n,m+2}}-u_{{n,m}} \right) }{u_{{n,m}} 
            \left( u_{{n,m+2}}{ \delta_{2}}+u_{{n,m+1}} \right) }}.
        \end{aligned}
        \label{eq:W21D4}
    \end{align}
    \label{eq:1D4int}
\end{subequations}

\begin{remark}
    \label{rem:1D4}
    The equation $_{1}D_{4}$ has an autonomous sub-case
    when $\delta_{1}=\delta_{2}=0$.
     {
    In this case the equations defining the first integrals
    become singular and the first integrals become
    simpler. If $\delta_{3}\neq0$ we have:
    \begin{subequations}
        \begin{align}
            W_{1}^{(0,0,\delta_{3})} = 
            \left( -1 \right)^{m}\frac{u_{n+1,m}-u_{n-1,m}}{u_{n,m}},
            \label{eq:W11D400d3}
            \\
            W_{2}^{(0,0,\delta_{3})} = 
            \left( -1 \right)^{n}\frac{u_{n,m+1}-u_{n,m-1}}{u_{n,m}}.
            \label{eq:W21D400d3}
        \end{align}
        \label{eq:1D400d3}
    \end{subequations}
    They are both non-autonomous, three-point, second order first integrals.
    Notice that, since this sub-case is such that
    the equation has the discrete 
    symmetry $n\leftrightarrow m$,
    the first integral $W_{2}^{(0,0,\delta_{3})}$ 
    can be obtained from $W_{1}^{(0,0,\delta_{3})}$ by using
    such transformation.

    Moreover we notice that the case $\delta_{1}=\delta_{2}=0$
    $\delta_{3}\neq0$
    is linked to the equation (4) with $b_{3}=1$ of \emph{List 3} in
    \cite{GarifullinYamilov2012}:
    \begin{equation}
        \hat{u}_{n+1,m+1}\hat{u}_{n,m} + 
        \hat{u}_{n+1,m}\hat{u}_{n,m+1}+1=0
        \label{eq:4list3}
    \end{equation}
    through the transformation
    $u_{n,m}=\sqrt{\delta_{3}}\hat{u}_{n,m}$.

    We finally notice that the sub-case where
    $\delta_{1}=\delta_{2}=\delta_{3}=0$ possesses two
    two-point, first order, non-autonomous first integrals:
    \begin{equation}
        W_{1}^{(0,0,0)} = 
        \left( -1 \right)^{m}\frac{u_{n+1,m}}{u_{n,m}},
        \quad
        W_{2}^{(0,0,0)} = 
        \left( -1 \right)^{n}\frac{u_{n,m+1}}{u_{n,m}}.
        \label{eq:1D4000}
    \end{equation}
    The resulting equation is linked to one of the
    linearizable and Darboux integrable cases presented
    in \cite{Hietarinta2005,GSL_Pavel}.
}

\end{remark}

\subsubsection{$_{2}D_{4}$ equation}

In the case of the $_{2}D_{4}$ equation given by \eqref{eq:2D4}, 
we have the following four-point, third order
first integrals:
\begin{subequations}
    \begin{align}
        W_{1} &
        \begin{aligned}[t]
            &={ \Fppp} \alpha {\frac { \left[
                \begin{gathered}
                    \left( u_{{n,m}}-u_{{n+2,m}}
                    -{ \delta_{1}}{ \delta_{2}}u_{{n-1,m}} \right)u_{n+1,m}^{2}
                    \\
                    +u_{{n+1,m}}u_{{n+2,m}}u_{{n-1,m}}
                    +{ \delta_{3}}u_{{n-1,m}}   
                \end{gathered}\right]
                }{ 
                \left( { \delta_{2}}u_{n+1,m}^{2}{ \delta_{1}}-
                { \delta_{3}}-u_{{n,m}}u_{{n+1,m}} \right) 
                u_{{n-1,m}}}}
            \\
            &-{ \Fpmm}\alpha {\frac {u_{{n+2,m}}u_{{n-1,m}}+ 
            \left( -u_{{n+2,m}}+u_{{n,m}} \right) 
            u_{{n+1,m}}+{ \delta_{3}}}{u_{{n-1,m}}u_{{n,m}}+{ \delta_{3}}}}
            \\
            &-{\Fmpm}\beta {\frac { \left( u_{{n+1,m}}-u_{{n-1,m}} \right)  
            \left( u_{{n+2,m}}-u_{{n,m}} \right) u_{{n,m}}}{u_{{n+2,m}} 
            \left( { \delta_{2}}{ \delta_{1}}{u_{{n,m}}}^{2}-u_{{n,m}}u_{{n+1,m}}-
            { \delta_{3}} \right) }}
            \\
            &+{ \Fmmp}\beta {\frac { \left( u_{{n+1,m}}-u_{{n-1,m}} \right)  
            \left( u_{{n+2,m}}-u_{{n,m}} \right) }{u_{{n+1,m}}u_{{n+2,m}}+{ \delta_{3}}}},
        \end{aligned}
        \label{eq:W12D4}
        \\
        W_{2} &
        \begin{aligned}[t]
            &={ \Fppp}\alpha \frac {\left(u_{{n,m+2}}-u_{{n,m}}\right) u_{{n,m-1}}
                +{ \delta_{1}}{ \delta_{3}}
                -{ \delta_{2}}u_{n,m+1}^{2}-u_{{n,m+1}}u_{{n,m+2}}
            }{{ \delta_{1}}{ \delta_{3}}-{ \delta_{2}}u_{n,m+1}^{2}-u_{{n,m}}u_{{n,m+1}}} 
            \\
            &-{ \Fpmm}\beta {\frac { \left( u_{{n,m+1}}-u_{{n,m-1}} \right)  
            \left( u_{{n,m+2}}-u_{{n,m}} \right) }{{ \delta_{1}}{ \delta_{3}}-{ \delta_{2}}u_{n,m}^{2}
            -u_{{n,m}}u_{{n,m+1}}}}
            \\
            &+{ \Fmpm}\alpha {\frac {u_{{n,m+2}}{ \delta_{2}}u_{{n,m}}+u_{{n,m-1}}u_{{n,m}}
            +u_{{n,m+1}}u_{{n,m+2}}-u_{{n,m}}u_{{n,m+1}}}{u_{{n,m+2}} 
            \left( { \delta_{2}}u_{{n,m}}+u_{{n,m-1}} \right) }} 
            \\
            &+{ \Fmmp}\beta{\frac { \left( u_{{n,m+1}}-u_{{n,m-1}} \right)  
            \left( u_{{n,m+2}}-u_{{n,m}} \right) }{ \left( { \delta_{2}}u_{{n,m+1}}+u_{{n,m+2}} \right) 
            u_{{n,m-1}}}}.
        \end{aligned}
        \label{eq:W22D4}
    \end{align}
    \label{eq:2D4int}
\end{subequations}

\begin{remark}
    \label{rem:2D4}
    The equation $_{2}D_{4}$ has an autonomous sub-case
    when $\delta_{1}=\delta_{2}=0$.
     {
    In this case the equations defining the first integrals
    become singular and the first integrals become
    simpler:
    \begin{equation}
        W_{1} ^{(0,0,\delta_{3})}= \left( -1 \right)^{m}
            \left( u_{n,m}u_{n+1,m}+\frac{\delta_{3}}{2} \right),
        \quad
        W_{2}^{(0,0,\delta_{3})} = \left( \frac{u_{n,m+1}}{u_{n,m-1}} \right)^{(-1)^{n}}.
        \label{eq:2D400d3}
    \end{equation}
    Therefore this particular sub-case possesses
    a two-point, first order integral $W_{1}$ and a three-point,
    second order integral $W_{2}$.
    Both integrals are non-autonomous despite the equation is autonomous.
    The limit $\delta_{3}\to0$ is in this case regular both
    in the first integrals and in the equations defining
    them.

    Finally we have that the case $\delta_{1}=\delta_{2}=0$
    corresponds
    to the equation (9) of \emph{List 4} in \cite{GarifullinYamilov2012}:
    \begin{equation}
        \hat{u}_{n,m}\hat{u}_{n+1,m}+\hat{u}_{n,m+1}\hat{u}_{n+1,m+1}+c_{4}=0,
        \label{eq:9list4}
    \end{equation}
    with the identification $u_{n,m}=\hat{u}_{n,m}$ and $c_{4}=\delta_{3}$.
    }

\end{remark}

\begin{remark}
    The first integrals of the \Hsechs~equations are rather
    peculiar.
     {Excluding the autonomous particular cases
        given in Remarks \ref{rem:1D4} and \ref{rem:2D4},
        we have that all the \Hsechs\ equations
        possess two different integrals in every
        direction. This is due to the presence
        of two arbitrary constants $\alpha$
        and $\beta$ in the expressions of the first
        integrals.}
    We believe that this reflects the fact that the $H^{6}$ equations
    on the lattice  {have two-periodic coefficients}
    in both directions.
\end{remark}

\section{Conclusions and outlook}
\label{sec:concl}

In this paper  {we have presented} a non-autonomous 
modified version of the algorithm developed in 
\cite{Habibullin2005,GarifullinYamilov2012,GarifullinYamilov2015}
 {to compute} the first integrals of two-dimensional
partial difference equations.
 {Applying this algorithm,}
we have showed that all the equations of the
trapezoidal $H^{4}$ and $H^{6}$ families, as given in
\cite{Boll2011,Boll2012a,Boll2012b},
 {possess first integrals in both directions, and
so have proved that these equations}
are Darboux integrable.
This result confirms the  {outcome} of the Algebraic
Entropy test presented in \cite{GSL_general}.

 {Furthermore the first integrals, even
those of higher order, can be used to find the
\emph{general solutions} of these equations.
Since this procedure is not trivial and
standard, we reserve its application
to a future work \cite{GSY_DarbouxII}.}
%
To be concrete,  {we will give an example
    on how this procedure can be carried out
in the case of}
the \tHeq{1} equation 
given by \eqref{eq:tH1e}, whose first integrals
are given by \eqref{eq:tH1eintegralslin}
and have been first presented in \cite{GSL_Pavel}.
 {We wish to solve the \tHeq{1} equation 
using both first integrals}.
We are going to construct those general solutions, 
 {slightly modifying the construction
    scheme from \cite{GarifullinYamilov2015}.}

Let us start from the integral $W_{1}$ \eqref{eq:W1tH1elin}.
This is a two-point, first order integral. 
 {This implies that the \tHeq{1} equation \eqref{eq:tH1e} can be
rewritten as the relation \eqref{eq:darb2} for the first integral $W_{1}$:} 
\begin{equation}
    \left( T_{m}-\Id \right)
    \left( \Fp{m}\frac{\alpha_{2}}{u_{n+1,m}-u_{n,m}}
    +\Fm{m}\frac{u_{n+1,m}-u_{n,m}}{1+\varepsilon^{2}u_{n,m}u_{n+1,m}} \right)
    =0.
    \label{eq:tH1easconslaw}
\end{equation}
From \eqref{eq:tH1easconslaw} we can derive the general solution
of \eqref{eq:tH1e} itself. In fact \eqref{eq:tH1easconslaw} implies:
\begin{equation}
    \Fp{m}\frac{\alpha_{2}}{u_{n+1,m}-u_{n,m}}
    +\Fm{m}\frac{u_{n+1,m}-u_{n,m}}{1+\varepsilon^{2}u_{n,m}u_{n+1,m}}
    =\lambda_{n},
    \label{eq:tH1esolstep1}
\end{equation}
where $\lambda_{n}$ is an arbitrary function of $n$. This is a first order
difference equation in the $n$-direction in which $m$ plays the role 
of a parameter. For this reason we can safely separate the two cases: 
$m$ even and $m$ odd.

\begin{description}
    \item[Case $m=2k$] In this case \eqref{eq:tH1esolstep1}
        is reduced to the linear equation
        \begin{equation}
            u_{n+1,2k} - u_{n,2k} = \frac{\alpha_{2}}{\lambda_{n}}
            \label{eq:tH1esolmev1}
        \end{equation}
        which has the solution
        \begin{equation}
            u_{n,2k} = \theta_{2k} + \omega_{n},
            \label{eq:tH1solmev2}
        \end{equation}
        where $\theta_{2k}$ is an arbitrary function and
        $\omega_{n}$ is the solution of the simple
        ordinary difference equation
        \begin{equation}
            \omega_{n+1} - \omega_{n} = \frac{\alpha_{2}}{\lambda_{n}},
            \quad
            \omega_{0} = 0.
            \label{eq:tH1esolmev3}
        \end{equation}
    \item[Case $m=2k+1$] In this case \eqref{eq:tH1esolstep1}
        is reduced to the discrete Riccati equation:
        \begin{equation}
            \lambda_{n}\varepsilon^{2}u_{n,2k+1}u_{n+1,2k+1}
            -u_{n+1,2k+1}+u_{n,2k+1}+\lambda_{n}
            =0.
            \label{eq:tH1esolmodd1}
        \end{equation}
        By using the M\"obius transformation
        \begin{equation}
            u_{n,2k+1} = \frac{\imath}{\varepsilon}\frac{1-v_{n,2k+1}}{1+v_{n,2k+1}},
            \label{eq:tH1esolmodd_moebius}
        \end{equation}
         {this equation can be recast into the linear equation}
        \begin{equation}
            \left( \imath+\epsilon\lambda_{{n}} \right) v_{{n+1,2k+1}}
            -\left( \imath-\epsilon\lambda_{{n}} \right) v_{{n,2k+1}}=0.
            \label{eq:tH1esolmodd2}
        \end{equation}
        If we introduce a new function $\kappa_{n}$, such that
        \begin{equation}
            \frac{\kappa_{n+1}}{\kappa_{n}} = \frac{\imath -\varepsilon \lambda_{n}}{%
                \imath + \varepsilon\lambda_{n}},
            \label{eq:kappandef}
        \end{equation}
        then we have that the general solution of  \eqref{eq:tH1esolmodd2} is
        expressed as:
        \begin{equation}
            v_{n,2k+1} = \kappa_{n}\theta_{2k+1}, 
            \label{eq:tH1esolmodd3}
        \end{equation}
				where $\theta_{2k+1}$ is an arbitrary function.
        Using \eqref{eq:tH1esolmodd_moebius} and \eqref{eq:kappandef} we then obtain:
        \begin{equation}
            u_{n,2k+1} = \frac{\imath}{\varepsilon}
                \frac{1-\kappa_{n}\theta_{2k+1}}{1+\kappa_{n}\theta_{2k+1}},
            \quad 
            \lambda_{n} = \frac{\imath}{\varepsilon}\frac{\kappa_{n}-\kappa_{n+1}}{\kappa_{n}+\kappa_{n+1}}.
            \label{eq:tH1esolmodd4}
        \end{equation}
So we have the general solution of \eqref{eq:tH1e} in the form:
\begin{equation}
    u_{n,m} = \Fp{m} \left( \theta_{m} + \omega_{n} \right)
    +\Fm{m}\frac{\imath}{\varepsilon}\frac{1-\kappa_{n}\theta_{m}}{%
        1+\kappa_{n}\theta_{m}},  
    \label{eq:tH1solfin}
\end{equation}
where $\theta_m$,$\kappa_{n}$ are arbitrary functions, $\omega_{n}$ is defined via $\lambda_{n}$ by \eqref{eq:tH1esolmev3}, 
and $\lambda_{n}$ is defined via $\kappa_{n}$ by
\eqref{eq:tH1esolmodd4}.
\end{description}

Now we pass to the integral in the direction $m$, namely, $W_{2}$ given by
\eqref{eq:W2tH1elin}. 
This case is more interesting, as now we are dealing
with a three-point, second order integral.
For this problem we can choose $\alpha=\beta=1$.
Our starting point is the relation  {\eqref{eq:darbeq2},
i.e. $W_{2}= \rho_{m}$,}
from which we can derive two different
equations, one for the even and one for the odd $m$. 
This gives \emph{a priori} a coupled system. 
However in this case, choosing $m=2k$ and $m=2k+1$, 
we obtain the following two equations:
\begin{subequations}
    \begin{align}
        1+\varepsilon^{2}u_{n,2k+1}u_{n,2k-1} &= \rho_{2k}
        \left({u_{n,2k+1}-u_{n,2k-1}}\right),
        \label{eq:tH1W2sys2k}
        \\
        u_{n,2k+2}-u_{n,2k} &= \rho_{2k+1}.
        \label{eq:tH1W2sys2kp}
    \end{align}
    \label{eq:tH1W2sys}
\end{subequations}
 {So the system consists of two \emph{uncoupled} equations.}

The first one \eqref{eq:tH1W2sys2k} is a discrete Riccati equation
which can be linearized through the non-autonomous M\"obius transformation:
\begin{equation}
    u_{n,2k-1} = \frac{1}{v_{n,k}} + \alpha_{k},
    \quad
    \rho_{2k} = \frac{1+\varepsilon^{2}\alpha_{k+1}\alpha_{k}}{\alpha_{k+1}-\alpha_{k}},
    \label{eq:mobW2}
\end{equation}
from which we obtain:
\begin{equation}
    \left( 1 + \varepsilon^{2} \alpha_{k+1}^{2} \right) v_{n,k+1}+ \varepsilon^{2}\alpha_{k+1}
    = 
    \left( 1 + \varepsilon^{2} \alpha_{k}^{2} \right) v_{n,k}+ \varepsilon^{2}\alpha_{k}.
    \label{eq:tH1W2sys2k2}
\end{equation}
This equation is equivalent to a total difference and therefore its solution is given 
by:
\begin{equation}
    v_{n,k} = \frac{\theta_{n} - \varepsilon^{2}\alpha_{k}}{1 + \varepsilon^{2} \alpha_{k}^{2} },
    \label{eq:tH1W2solv}
\end{equation}
with an arbitrary function $\theta_{n}$.
Putting $\alpha_{k} = \kappa_{2k-1}$, we obtain the solution for 
$u_{n,2k-1}$:
\begin{equation}
    u_{n,2k-1} = \frac{1+\kappa_{2k-1} \theta_{n}}{\theta_{n} - \varepsilon^{2} \kappa_{2k-1}}.
    \label{eq:tH1W2soluev}
\end{equation}

The second equation is just a linear ordinary difference  equation
which can be written as a total difference, performing the substitution 
$\rho_{2k+1} = \kappa_{2k+2}-\kappa_{2k}$, and we get:
\begin{equation}
    u_{n,2k} = \omega_{n} + \kappa_{2k}.
    \label{eq:tH1W2soluodd}
\end{equation}

The resulting solution reads:
\begin{equation}
    u_{n,m} = \Fp{m} \left( \omega_{n} + \kappa_{m} \right)
    +\Fm{m} \frac{1+\kappa_{m} \theta_{n}}{\theta_{n} - \varepsilon^{2} \kappa_{m}}.
    \label{eq:tH1W2solu}
\end{equation}
This solution depends on \emph{three} arbitrary functions.
This is because we started from a second order first integral,
which is just a consequence of the discrete equation.
This means
that there must be a relation between $\theta_{n}$ and $\omega_{n}$.
This relation can be retrieved by inserting \eqref{eq:tH1W2solu} into
\eqref{eq:tH1e}.
As a result we obtain the following definition for $\omega_{n}$:
\begin{equation}
    \omega_{n}-\omega_{n+1} = \alpha_{2} 
    \frac{\varepsilon^{2}+\theta_{n}\theta_{n+1}}{\theta_{n+1}-\theta_{n}},
    \label{eq:tH1omthetarel}
\end{equation}
which gives us the final expression for the solution of \eqref{eq:tH1e} 
up to the discrete integration given by \eqref{eq:tH1omthetarel}.
The general solutions obtained from different first integrals are 
the same in the sense that one of them can easily be transformed 
into the other one. 

As a final remark we note that it has been proved in \cite{AdlerStartsev1999} 
that Darboux integrable
systems possess generalized symmetries depending on arbitrary functions
of the first integrals. However, in case of the
trapezoidal $H^{4}$ and $H^{6}$, the explicit form of symmetries
depending on arbitrary functions is known only for the \tHeq{1} equation
\eqref{eq:tH1e} \cite{GSL_symmetries,GSL_Gallipoli15,GSL_Pavel}.
This poses the challenging problem of finding the explicit form of such
generalized symmetries. These symmetries will be
highly nontrivial, especially in the case
of the \tHeq{2} and \tHeq{3} equations \eqref{eq:trapezoidalH4},
where the order of the first integrals is particularly high.

\section*{Acknowledgments}

GG is supported by  INFN  IS-CSN4 \emph{Mathematical Methods of
Nonlinear Physics}.
RIY gratefully acknowledges financial support
from a Russian Science Foundation grant (project 15-11-20007).

\bibliographystyle{unsrt}
\bibliography{bibliography}

\end{document}